\documentclass[5p]{elsarticle}

\usepackage{hyperref}
\hypersetup{
    colorlinks  = false,
    pdfborder   = {0 0 0},
    pdftitle    = {Classical and Machine Learning Methods for Event Reconstruction in NeuLAND},
    pdfauthor   = {Jan Mayer},
}

\usepackage{amsmath}
\usepackage{siunitx}
\usepackage{booktabs}
\usepackage{multirow}
\usepackage{tikz, flowchart}
\usetikzlibrary{arrows.meta}
\usetikzlibrary{shapes,arrows.meta,chains,positioning,calc,fit,shadows}
\usepackage[capitalise]{cleveref}
\usepackage{isotope}
\usepackage{csquotes}
\usepackage{setspace}

\journal{Nuclear Instruments and Methods in Physics Research Section A}
\biboptions{sort&compress}

\begin{document}
\begin{frontmatter}

\title{Classical and Machine Learning Methods for Event Reconstruction in NeuLAND}

\author[ikp]{Jan Mayer\corref{corres}}
\ead{jan.mayer@ikp.uni-koeln.de}
\author[gsi]{Konstanze Boretzky}
\author[kvi]{Christiaan Douma}
\author[ikp]{Elena Hoemann}
\author[ikp]{Andreas Zilges}
\author{for the R³B collaboration}

\cortext[corres]{Corresponding author}
\address[ikp]{Institute for Nuclear Physics, University of Cologne, Germany}
\address[gsi]{GSI Helmholtzzentrum für Schwerionenforschung GmbH, Darmstadt, Germany}
\address[kvi]{KVI-CART, University of Groningen, Netherlands}

\begin{abstract}
NeuLAND, the New Large Area Neutron Detector, is a key component to investigate the origin of matter in the universe with experimental nuclear physics.
It is a core component of the Reactions with Relativistic Radioactive Beams setup at the Facility for Antiproton and Ion Research, Germany.
Neutrons emitted from these reactions create a wide range of patterns in NeuLAND.
From these patterns, the number of neutrons (multiplicity) and their first interaction points must be reconstructed to determine the neutrons' four-momenta.
In this paper, we detail the challenges involved in this reconstruction and present a range of possible solutions.
Scikit-Learn classification models and simple Keras-based neural networks were trained on a wide range of input-scaler combinations and compared to classical models.
While the improvement in multiplicity reconstruction is limited due to the overlap between features, the machine learning methods achieve a significantly better first interaction point selection, which directly improves the resolution of physical quantities.

\hspace*{0.2mm}

\noindent Published in Nuclear Instruments and Methods in Physics Research Section A 1013, 165666 (2021).

\noindent \href{https://doi.org/10.1016/j.nima.2021.165666}{DOI: 10.1016/j.nima.2021.165666} © 2021. This manuscript version is made available under the CC-BY-NC-ND 4.0 license \url{http://creativecommons.org/licenses/by-nc-nd/4.0/}.
\end{abstract}

\end{frontmatter}

\section{Fundamental Science with NeuLAND}
One goal of the science program at the Facility for Antiproton and Ion Research (FAIR) \cite{FAIR, Aumann2007}, the \emph{Universe in the Laboratory}, is to unravel the origin and properties of matter in the universe.
It is unclear which stellar explosion conditions are responsible for the production of the different isotopes, especially those far away from stability.
At the Reactions with Relativistic Radioactive Beams (R\textsuperscript{3}B) experiment at FAIR, these exotic nuclei can be studied using various reaction types and analysis methods.
A radioactive ion beam is produced by an accelerator complex, where a relativistic primary beam hits a primary target and fragments into a wide range of different isotopes.
The FRagment Separator (Super-FRS) \cite{Geissel2003} selects the desired nuclei and routes them to the R\textsuperscript{3}B experiment, where they hit the secondary target and the reaction of interest can be studied.
The target is surrounded by detectors for light particles and $\gamma$-rays.
The large superconducting magnet GLAD deflects charged particles from the original path of the beam.
Numerous tracking detectors gather information on the flight paths of the ejectiles to determine their type and energy.
Neutrons are not affected by the magnetic field and can be detected by the New Large Area Neutron Detector NeuLAND located downstream at zero degrees.

Nuclear properties can be inferred from the gathered data in several ways.
For example, the invariant mass method is often used with this type of setup \cite{Aumann2005, Baumann2012}.
The relative energy $E_{rel}$
\begin{equation}
E_{rel} = \left( \left|\sum_i \mathbf{P}_i\right| - \sum_i m_i \right) c^2, \label{e:erel}
\end{equation}
is a measure for the decay energy of unbound states.
The four-momenta $\mathbf{P}_i$ of all participating particles must be known with high precision.

\begin{figure}[htb]
\begin{tikzpicture}[>=Latex, font=\sffamily, x=5.57mm, y=5.57mm]
\node [anchor=north west,inner sep=0] (img) at (0,0) {\includegraphics[width=\columnwidth, trim={1cm 3cm 1cm 1cm}, clip]{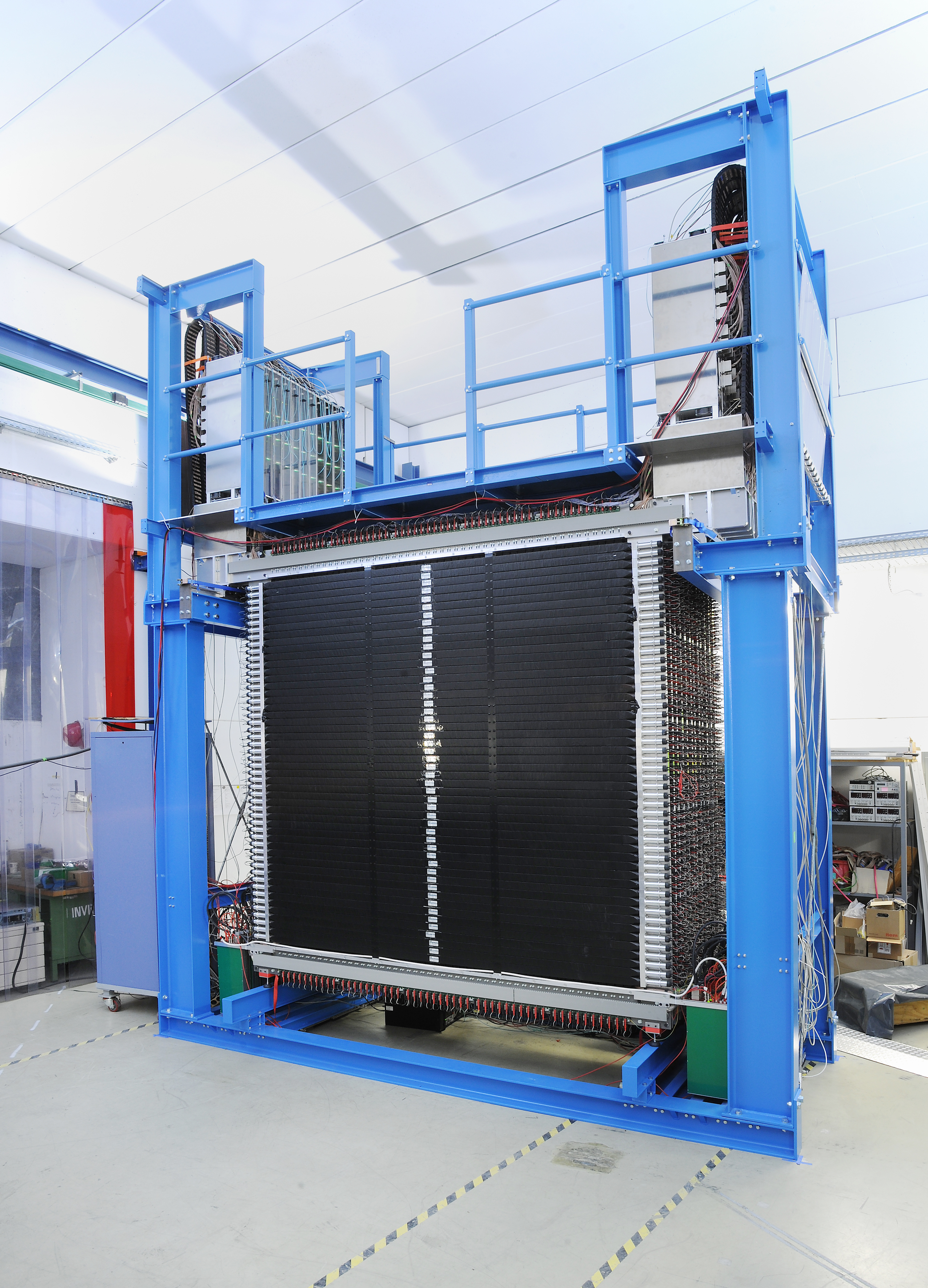}};
\tikzset{annotation/.style={draw, fill=white!20, opacity=.8,text opacity=1}}
\node [annotation, anchor=west] at (3.5,-7.1) (el) {Readout Electronics};
\node [annotation, anchor=west] at (9,-20) (hv) {High Voltage Supply};
\draw [thick, <->, color=white] (4.3,-10.2) -- (4.45,-16.92) node [midway,right] {2.5\,m};
\draw [thick, <->, color=white] (4.45,-16.92) -- (11.1,-17.7) node [midway,above,xshift=0.5cm] {2.5\,m};
\end{tikzpicture}
\caption{
NeuLAND, the New Large Area Neutron Detector, in the experimental hall at GSI/FAIR.
Plastic scintillator bars with a thickness of 5\,cm and a length of 2.7\,m are wrapped in black tape and arranged into detector modules.
Each double plane contains 100 bars, 50 in vertical and 50 in horizontal orientation, creating a face area of 2.5\,m times 2.5\,m.
At the ends of each bar, photomultipliers are attached to detect the light created in the scintillator.
The high-voltage required for the photomultipliers is provided by a distribution system at the bottom of the detector, while the signals are digitized in  electronics modules at the top.
\label{fig:neuland}}
\end{figure}
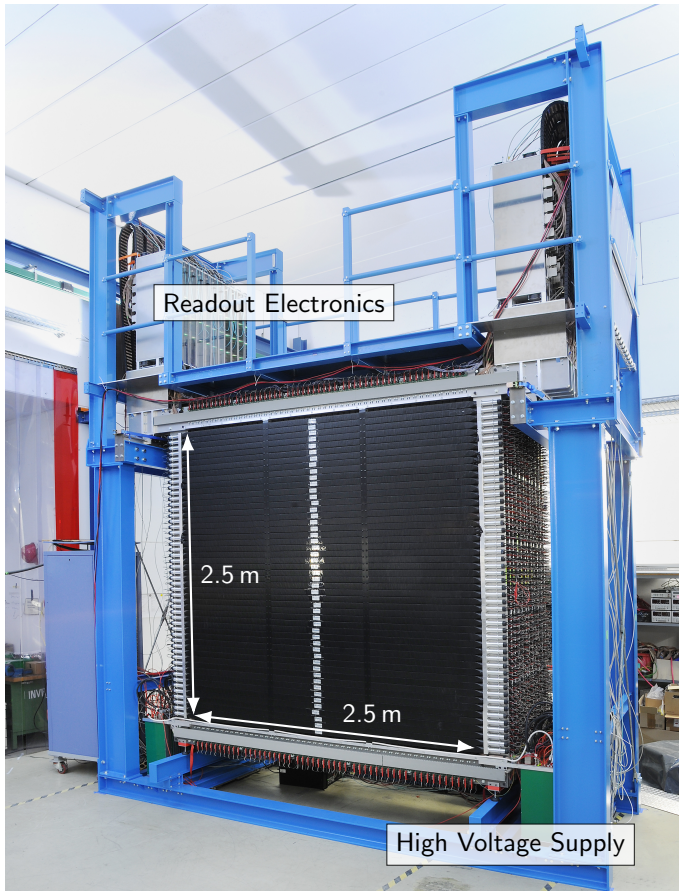

NeuLAND \cite{NeuLANDpaper, NeulandTDR} is dedicated to the simultaneous detection of up to five neutrons with kinetic energies up to \SI{1}{\GeV}.
Here, the four-momentum of each neutron is calculated from the time and position of its first interaction in the detector.
NeuLAND is built out of organic scintillator bars with a square profile of \SI{5}{\cm} by \SI{5}{\cm} and a length of \SI{270}{\cm}, including a \SI{10}{\cm} conical taper at both ends to which photomultiplier tubes (PMTs) with a diameter of \SI{2.54}{\cm} are connected.
The bars are arranged to double planes with 50 horizontal bars in front of 50 vertical bars.
This creates a face area of \SI{250}{\cm} by \SI{250}{\cm}, see \cref{fig:neuland}.
Each double plane is an independent unit with its own electronics and voltage supply.
These double planes can be arranged in different detector configurations if needed.
For most experiments, however, the double planes are placed directly behind each other to form a single, large detector, as the neutron detection capability increases with detector depth.
In its final configuration, the detector consists of 30 double planes with a total of 3000 scintillator bars and 6000 data channels.
At the time of writing, experiments have been performed with up to 12 double planes.

In this paper, we discuss the main challenges and several ideas to reconstruct the neutron multiplicity and the first neutron interaction points for the experiment as a whole and for each individual recorded reaction (event-by-event).
First, the main goals and challenges are presented, followed by a detailed discussion of \emph{usable properties}, called \emph{features} in machine learning terms.
We present different approaches from classic cuts over elementary Bayesian statistics to machine learning with and without neural networks and compare their performance for a specific test case.

\section{Goals and Challenges}
NeuLAND can deliver several types of information according to the needs of the experiment:
A binary detection condition for the trigger, the overall multiplicity distribution, the multiplicity on an event-by-event basis, and the neutron interaction points on an event-by-event basis are examples for such quantities in ascending order of reconstruction difficulty.

To reduce the amount of data recorded in the experiment, trigger conditions are applied to the data acquisition.
The trigger from NeuLAND is based on the number of channels with signals from PMTs above a threshold.
Typically, two signals above the threshold are demanded for the NeuLAND trigger.

\begin{figure*}
\centering
\includegraphics[width=0.32\textwidth]{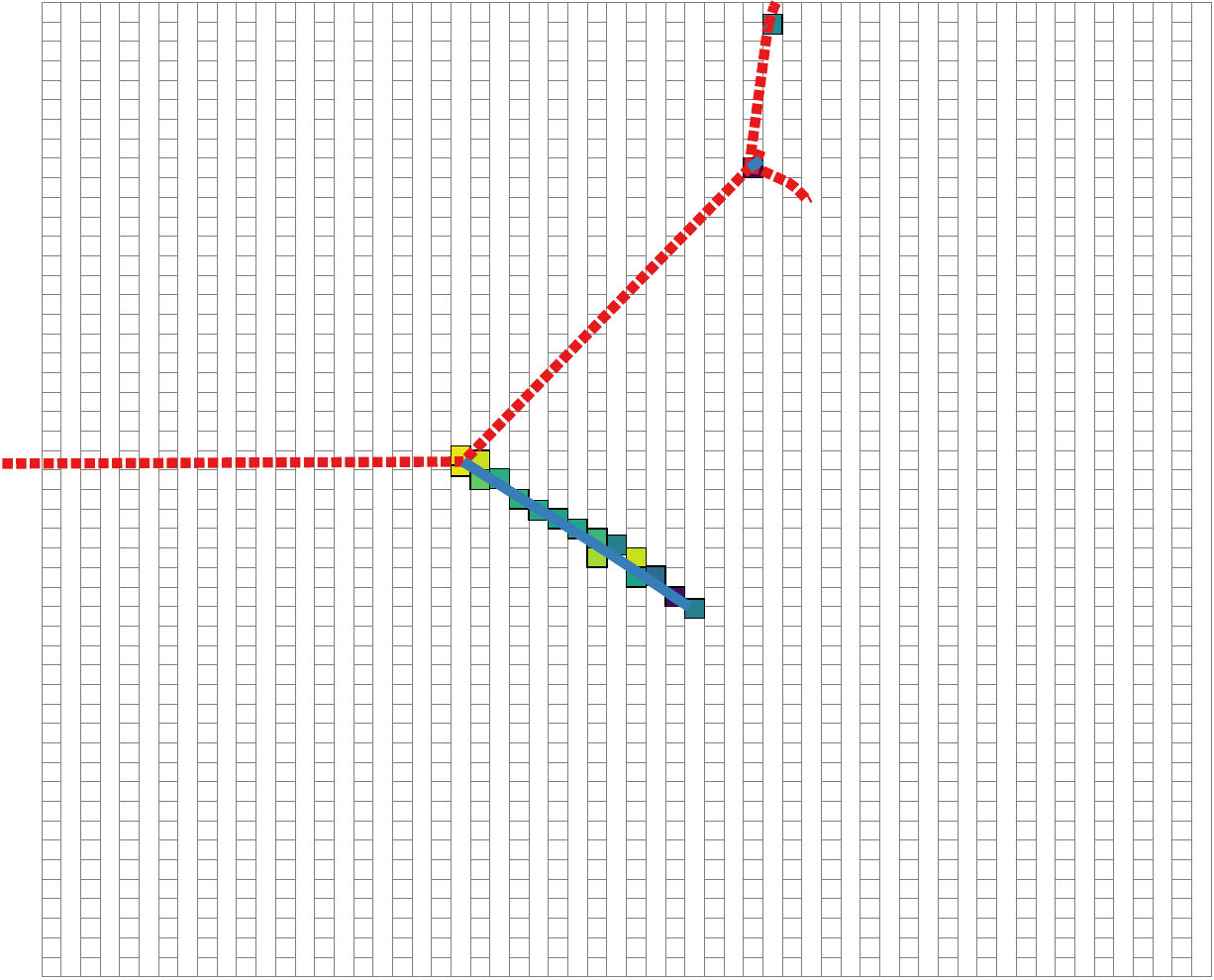}
\includegraphics[width=0.32\textwidth]{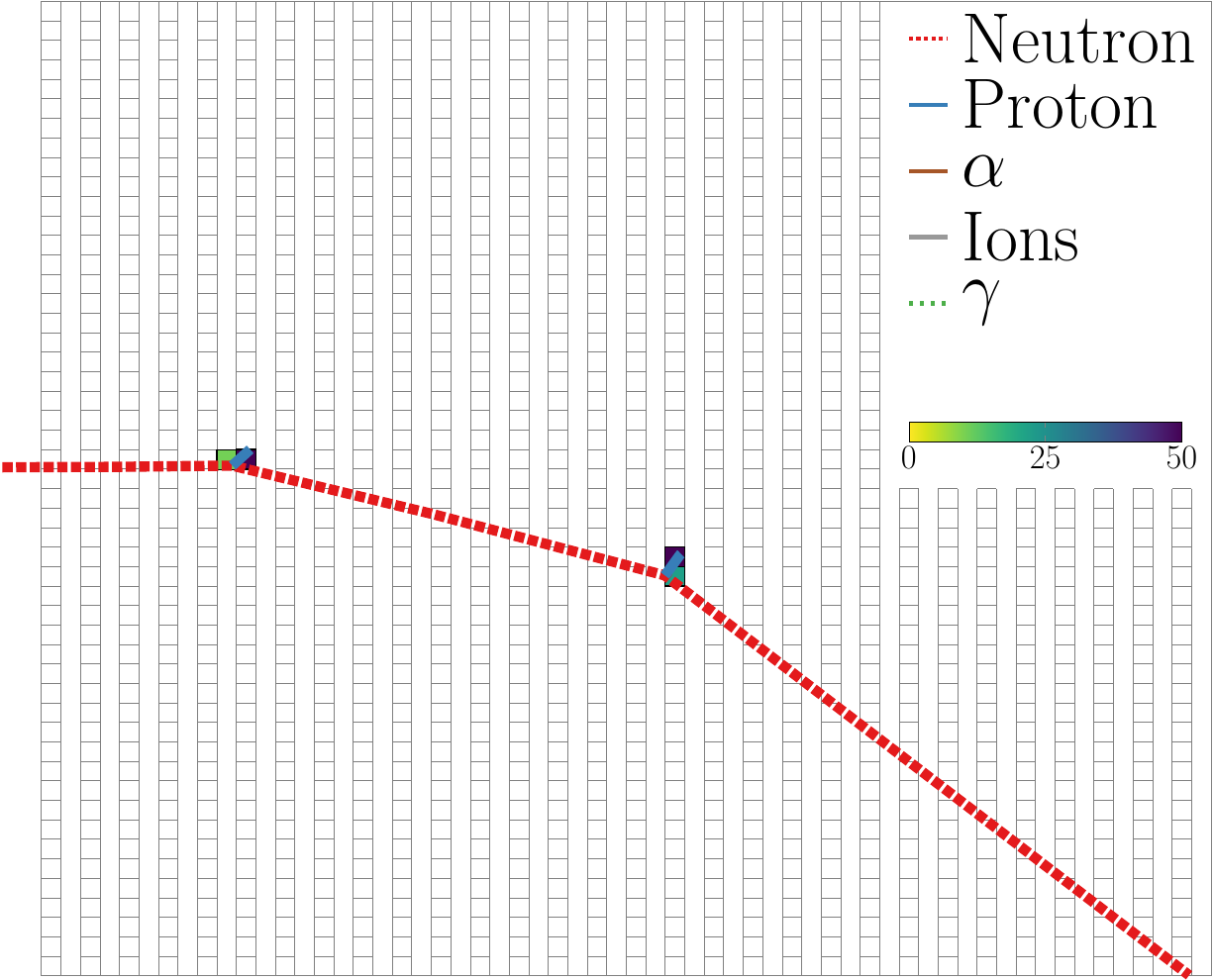}
\includegraphics[width=0.32\textwidth]{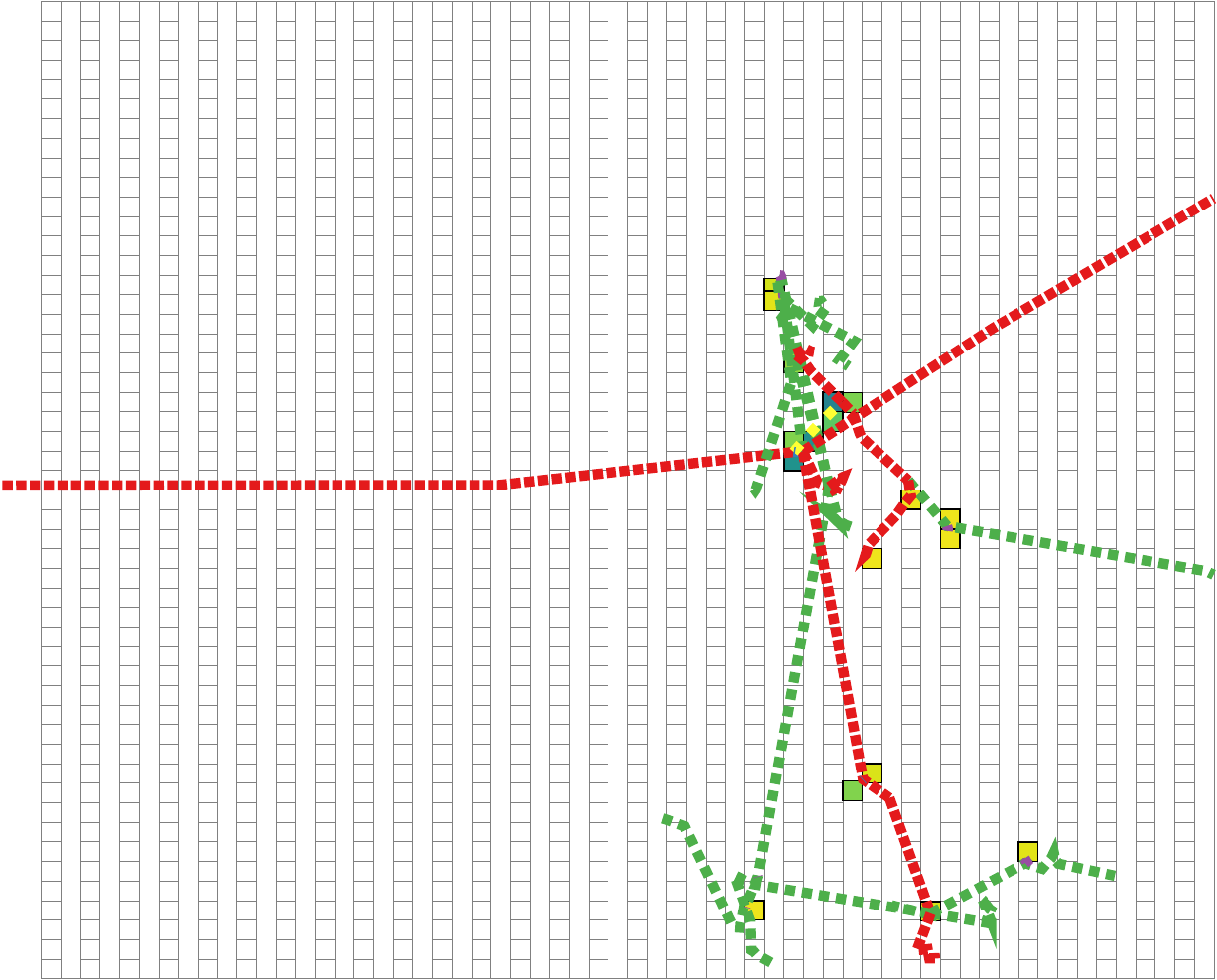}
\caption{
Side view of hit patterns in NeuLAND, created by the interaction of one neutron with a kinetic energy of \SI{600}{\MeV}.
The grid represents the alternating horizontal and vertical scintillator bars and the hit color represents the deposited energy in MeV.
Neutrons can interact at different positions and via different reactions, which results in different patterns: On the left, the incoming neutron scatters on a proton, which deposits its energy in a clearly defined track with the highest energy deposition at the end. In the middle, energy is deposited close to the first interaction point. On the right, energy depositions by secondary particles create many hits distributed over a wide area.
For more than one incoming neutron, these patterns overlap.
\label{f:hitpatterns}}
\end{figure*}

Deducing the number of neutrons that have been emitted is one of the primary deliverables of NeuLAND.
While detecting a single neutron is straightforward, differentiating between multiple neutrons is challenging.
Of the neutrons impinging on the detector, only some or even none might react, depending on the detector depth and the neutron energy.
The neutrons that do react in the detector do not always exhibit an easily recognizable pattern as a charged particle would.
Instead, a large variety of reactions can occur, with a wide range of deposited energy and distance between reactions within the detector, see \cref{f:hitpatterns}.
For multiple neutrons, these statistical properties are folded, and by adding conditions on a singular quantity (\emph{cutting}), like the total deposited energy, the neutron multiplicity cannot be extracted unambiguously.
The overall multiplicity distributions can be approximated as a linear combination of these individual distributions; however, this is rarely useful.
Typically, the multiplicity must be known for each event.
We have investigated cuts in a multidimensional space, a probabilistic approach with Bayesian statistics, and several types of machine learning and neural network approaches, see \cref{s:multiplicity}.

This high variance in interaction patterns also impedes finding the first interaction points of the primary neutrons, the second main deliverable.
From these coordinates, the four-momenta can be determined with high precision.
Finding the correct interaction points is essential for the invariant mass and other analysis methods \cite{Baumann2012}.

Multiplicity and first interaction points are related but distinct: Even if the multiplicity is given, the correct primary interaction points may not be obtainable. If no multiplicity limit is given, too many interaction points might be classified as primary.

The correct result of the reconstruction process is required beforehand for training and evaluation.
In machine learning terms, this is often called the \emph{label}.
For NeuLAND, we define different stages in the simulation process that can take the role of the \emph{label}:
\begin{description}
\item[Primary Neutrons]{are the fast neutrons emitted in the reaction studied in the experiment directed towards the detector.
Their number (multiplicity, $N_{PN}$) and their kinetic energy are the main quantities NeuLAND should deliver. Their number is also sometimes called the \emph{generated multiplicity}.}
\item[Primary Points]{are the exact positions where the primary neutrons interacted first during the Monte-Carlo transport. Not all incoming neutrons will react, thus the number of points might be smaller ($N_{PP} \le N_{PN}$).}
\item[Primary Hits]{are the positions where the first interaction is detected after the digitization of the energy depositions. Not all primary interactions will be detectable ($N_{PH} \le N_{PP}$) and their position might differ from the actual interaction point.}
\item[Primary Clusters]{are the groups of hits (see \cref{s:digiclus}) that include primary hits. Primary clusters are easier to identify than primary hits.}
\end{description}

The number of primary neutrons is the desired multiplicity quantity.
As neutrons might pass through the detector without leaving any trace at all, there can be fewer primary hits than primary neutrons --- especially at limited detector depth.
Here, we test the multiplicity reconstruction models with both quantities.

\section{Data generation and processing}\label{s:degen}
For the calibration of the reconstruction algorithm (\emph{training} the \emph{model} in machine learning terms), a large amount of simulated data is required.
Currently, the trained models are also evaluated on simulated data, either from a train-test-split of the same scenario or different scenarios.
A scenario represents one detector configuration, characterized predominantly by the number of double planes and distance, and one reaction type, characterized by beam energy, fragments, and angular spread.

At later stages, we will apply models trained with simulated data to experimental data.
We expect that models will have to be retrained for every experimental scenario, as changes in, e.g., the primary neutron energy or emission angle distributions, result in different hit distributions.

NeuLAND is implemented within R3BRoot, the software package for the R\textsuperscript{3}B experiment \cite{Bertini2011}.
It is based on FAIRRoot and handles raw experimental data processing, simulation, raw simulation data processing, data reconstruction, and data analysis \cite{Al-Turany2012}.
In turn, FAIRRoot is based on ROOT, an omnipresent software package in nuclear and particle physics \cite{Antcheva2009}.

We have tested reconstruction solutions integrated into R3BRoot, either directly as compiled code, indirectly by calling Python functions from compiled code, or communicating with other services with messaging protocols.
Other solutions use external scripts connected via non-ROOT storage such that (GPU-accelerated) systems without ROOT can be used.
Several different storage systems and formats are available, for example, the hierarchical data format HDF5 \cite{hdf5}, serialized (\emph{pickled}) tabular data like Pandas DataFrames \cite{mckinney-proc-scipy-2010, reback2020pandas}, serialized structured data formats like Google's Protocol Buffers \cite{protobuf}, or, what we prefer due to speed and file size, the column-oriented Apache Parquet\cite{parquet} format.

\subsection{Data flow in experiments and simulation}
\begin{figure}
\centering
\resizebox{\columnwidth}{!}{
\begin{tikzpicture}[node distance = 0.5cm and 0.75cm, minimum height=0.66cm, >=Latex, font=\sffamily]
\node (hits) [draw, storage] {Detector Hits};

\node (digitization) [draw, process, left = of hits, yshift = +0.75cm] {Digitization};
\node (simulation)   [draw, process, above = of digitization] {Simulation};
\draw[->] (simulation) -- (digitization);
\draw[->] (digitization) -- (hits);

\node (calibration)  [draw, process, left = of hits, yshift = -0.75cm] {Calibration};
\node (experiment)   [draw, process, below = of calibration] {Experiment};
\draw[->] (experiment) -- (calibration);
\draw[->] (calibration) -- (hits);

\node (reconstruction) [draw, process, right = of hits] {Reconstruction};
\node (physics) [draw, storage, right = of reconstruction] {Physics Data};
\draw[->] (hits) -- (reconstruction);
\draw[->] (reconstruction) -- (physics);

\node (comparison) [draw, process, above = of reconstruction, yshift = +0.75cm] {Optimization};
\draw[->] (simulation) -- (comparison);
\draw[->] (physics)  |- (comparison);
\draw[->] (comparison) -- (reconstruction);
\end{tikzpicture}
}
\caption[Generalized data flow scheme of R3BRoot.]{
Generalized data flow scheme of R3BRoot.
Raw experimental and simulated data are processed to detector hits which are then reconstructed to extract the physical data of interest. As the correct result is known in simulations, the effectiveness of the reconstruction stage can be evaluated and optimized.
\label{f:recoconcept}}
\end{figure}
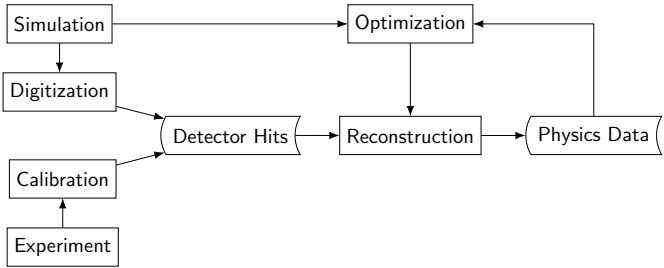

Experimental and simulated data are evaluated with the same routines.
After calibration of the experimental and digitization of the simulated data, the properties of both should match.
For NeuLAND, the main parameters are energy, time, and position in each bar, which together form a \emph{hit}.
Reconstruction algorithms can then be applied to a set of hits.
For simulated data, the result can be compared with the input, and thus the full data analysis stack can be constructed and tested before it is applied to experimental data, see \cref{f:recoconcept}.

During and after the experiment, raw data from the detectors is fed into the system through an \emph{unpacker}, which translates the raw binary formats into data storage classes.
From here on, R3BRoot passes the data through mapping and calibration stages.
Data is grouped in so-called events, where all hits in all detectors recorded in a short time window should belong to a single nuclear reaction.
A single experiment might create billions of events with storage requirements in the order of TB, with varying degrees of quality.
The resulting hit-level data with physical quantities is then reconstructed to extract the properties of the detected neutrons for further use in the actual nuclear physics analysis.

Alternatively, data can be created with simulations.
Monte-Carlo transport codes, a well-established example is \textsc{Geant4} \cite{Agostinelli2003}, implement the physics of particles passing through and reacting with matter.
These particle transport codes take a fixed virtual representation of the detector geometry plus the initial particle configuration and then randomly process possible reactions as well as resulting reaction chains in the material.
Each random sample is an event, and sampling ten thousand to several hundred thousand events is needed to gain statistical relevance.
The particle transport codes provide position, time, and deposited energy for interactions in the active detector material in small steps.
This simulated data needs to be processed to be equivalent to measured experimental data, which includes processes in the detector material like light generation, the response of the PMTs and data acquisition, and, not to forget, the calibration process, see \cref{s:digiclus}.
In the FairRoot framework, this transformation is called \emph{digitization} (even though no analog signals are involved).

We have verified the older NeuLAND TACQuila electronics with calibration data obtained with four double planes in an experiment at RIKEN in Japan and found an acceptable agreement \cite{Kahlbow2018}.
We do not expect major algorithm-breaking differences for the new TAMEX-based electronics.
A calibration experiment with the new electronics and more double planes will be performed at FAIR.

\subsection{Simulation}
For the results presented here, simulations were carried out with R3BRoot based on FairSoft \texttt{jun19p2} \cite{fairsoft} (a software bundle with \textsc{Geant4} \texttt{10.5.1.} \cite{Agostinelli2003}) and FairRoot \texttt{v18.2.1} \cite{Al-Turany2012}.

The primary particles for both the training and test data set were generated for the decay of \isotope[132]{Sn} into one to six primary neutrons and the remaining heavy particle \isotope[132-n]{Sn}.
All random number generators operated with different seeds for each simulation run.
Note that the usage of similar primary trajectories for training and testing might not produce a universally applicable model, which is discussed later.

For the geometry, we used a simplified version of the R\textsuperscript{3}B setup which only includes relevant parts.
This includes the magnetic field (to bend the heavy fragments away) without the magnet itself, the steel vacuum exit window of the magnet, the air between the vacuum window and NeuLAND, and NeuLAND itself.
Models were trained and tested on a split of simulated data for 12 and 30 double planes at a distance of \SI{15}{m} from the target, a neutron energy of \SI{600}{MeV}, and an $E_{rel}$ (see \cref{e:erel}) of \SI{500}{keV}.

In the final R3B configuration at FAIR, NeuLAND can be moved up to \SI{34}{\m} from the target.
This increase in flight distance leads to an improvement of the time-of-flight energy resolution, which might improve the effectiveness of the models which use it as a feature, see \cref{f:clusterfeatures}.

The interactions simulated in the Monte-Carlo codes are based on several different models and implementations, assembled into so-called \emph{physics lists}.
In other investigations, it was found that they introduce some uncertainty \cite{Douma2021}.
In this paper, we restrict the discussion to the \texttt{QGSP\_INCLXX\_HP} physics lists \cite{inclxx,PhysRevC.96.054602}.

\subsection{Digitizing and Clustering}\label{s:digiclus}
The simulated energy depositions are converted to hits with the same properties as real data.

Individual hits in NeuLAND are grouped in \emph{clusters}.
A single interaction of a high-energy neutron can result in multiple associated hits - the most iconic case are Bragg-Tracks from secondary protons, as shown in \cref{f:hitpatterns}.
Confining these physically related energy depositions in one entity simplifies the reconstruction process, as it reduces the number of first interaction candidates.
The properties as well as the number of clusters are also useful for reconstruction purposes, as discussed in \cref{s:clusterfeatures}.

A limit on the distance between hits can represent the physical relation.
This clustering condition is a trade-off between ensuring the inclusion of correlated hits and avoiding the inclusion of uncorrelated hits. We found it sufficient to include only directly adjacent scintillator bars, expressed as a multiple of the bar width $d_\text{bar} = \SI{5}{\cm}$ within a short time window:
\begin{equation}
\left|\Delta x\right| \leq 1.5 \cdot d_\text{bar} \land
\left|\Delta y\right| \leq 1.5 \cdot d_\text{bar} \land
\left|\Delta z\right| \leq 1.5 \cdot d_\text{bar} \land
\left|\Delta t\right| \leq \SI{1}{\ns}
\end{equation}

To ensure that all hits are assigned to the right cluster, we have implemented an algorithm dubbed \emph{handshake-chain-clustering}.
Here, the unsorted list of hits is partitioned into a clustered and unclustered part.
At the start, the first hit in the list is the reference hit and all other hits that fulfill the clustering condition (those who can \emph{shake hands} with the reference cluster) are moved to the clustered part.
Then the next hit in the clustered part is used as a reference and so on until the cluster is no longer growing and its end has been reached.
Thus, all hits shaking hands with other hits are clustered together.

The algorithm is implemented as a \texttt{C++} template; thus, any list of elements can be clustered if a binary clustering condition can be formulated.
It does not scale well with the number of elements to clusters, but as events seldom contain more than a hundred hits, see \cref{f:eventfeatures}, this is not an issue in this application.

\subsection{Primary points and hits}
Primary points and primary hits are obtained by tracing their origin back through the Monte-Carlo steps.

First, the primary neutrons are identified.
The simulation can either generate neutrons directly or generate other particles which in turn can produce neutrons with nuclear reactions in a target like a beam would in an experiment.
In the case that the incoming neutrons are created by the particle gun, e.g., when using precomputed input files, they can be immediately identified by their ID.
If the neutrons originate from a particle hitting a target, they must be traced back to this process and filtered based on their four-vectors such that only neutrons impinging on NeuLAND with enough kinetic energy are considered.

Second, the primary interaction points are identified.
Each energy deposition (point) in NeuLAND is associated with a primary neutron by stepping back through the Monte Carlo tracks, starting with the track that created the specific energy deposition.
Then the first point in time is taken for each primary track.

Primary hits are identified similarly.
Each point is associated with a hit, if possible, by comparing the ID of their detector element.
Then the first hit in time is taken for each primary track from the hits that originate from the points of this primary track using the point-to-track relation from above.

The clusters can then be split into groups of primary and secondary clusters by checking for the inclusion of a primary hit.

\section{Reaction Probability}\label{s:reacprob}
Neutrons, in contrast to charged particles, can only interact with matter via nuclear reactions, including scattering.
These randomly occurring processes can be described using probability theory.

For $N_{PN}$ incoming neutrons, the probability $P$ for $N_{PH}$ interactions is expected to be a binomial distribution
\begin{equation}
P = \binom{N_{PN}}{N_{PH}} \cdot p(n_\text{DP})^{N_{PH}} \cdot (1-p(n_\text{DP}))^{N_{PN} - N_{PH}},\label{eq:probbinom}
\end{equation}
where the interaction probability $p(n_\text{DP})$ for a detector depth of $n_\text{DP}$ double planes is given by
\begin{equation}
p(n_\text{DP}) = 1 - (1-p_\text{DP})^{n_\text{DP}}.
\end{equation}

\begin{figure}
\includegraphics[width=\columnwidth]{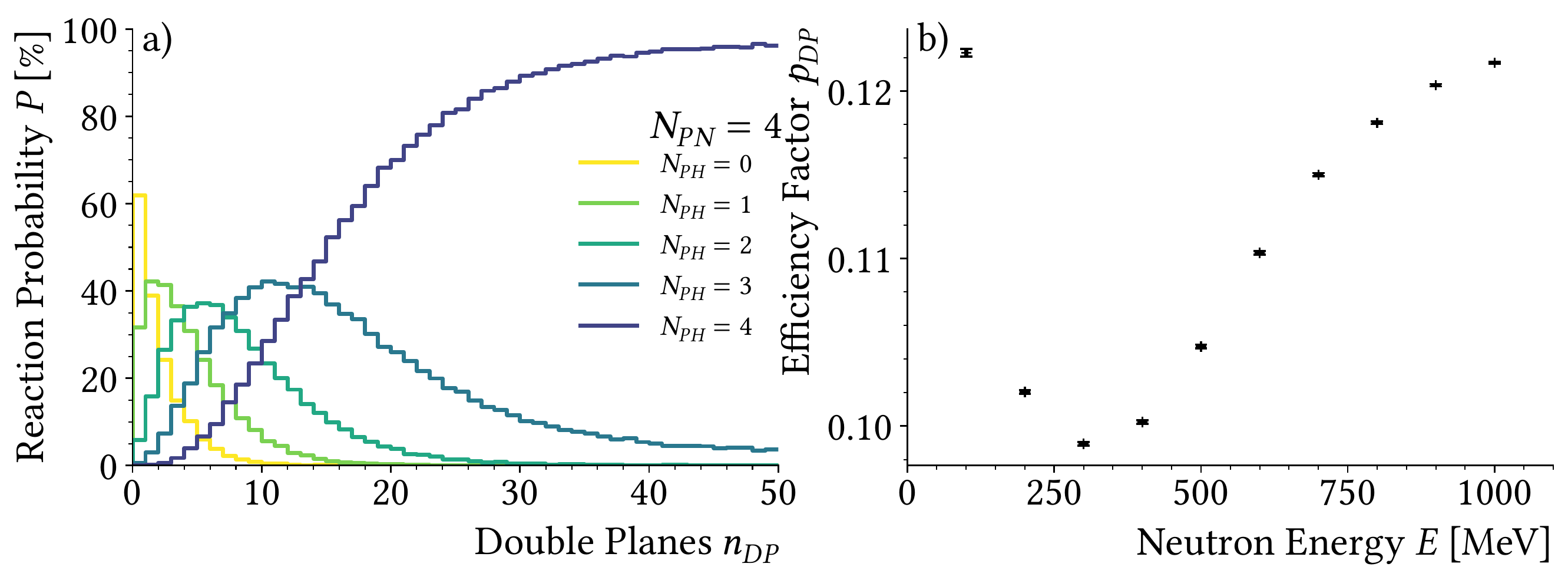}
\caption[Neutron reaction probability distributions as a function of the number of double planes for \SI{600}{\MeV}.]{
a) Neutron reaction probability distributions for four impinging neutrons as a function of the number of double planes for \SI{600}{\MeV} and b) the double plane efficiency factor $p_{DP}$ as a function of the neutron energy.
For $N_{PN}$ incoming neutrons, the number of reacted neutrons $N_{PH}$ is extracted from Monte Carlo data for each number of double planes $n_{DP}$.
These simulated probabilities follow the expected binomial distribution (\cref{eq:probbinom}), which can be fitted with the free parameter $p_{DP}$.
The intrinsic maximal achievable performance of NeuLAND, i.e., where all incoming neutrons undergo a reaction in the detector volume ($N_{PN} = N_{PH}$), strongly rises when increasing the detector depth from 10 to 20 double planes, while at the same time suppressing unwanted channels ($N_{PH} < N_{PN}$).
\label{fig:probbinom}}
\end{figure}

The \emph{double plane efficiency factor}, i.e., the probability $p_{DP}$ for a primary neutron interaction in one NeuLAND double plane, only depends on the neutron energy and can be determined by simulations.
A fit of \cref{eq:probbinom} to distributions from one to five incoming neutrons for one to 50 double planes with $p_{DP}$ as the only free parameter matches the simulated distributions very well.
In \cref{fig:probbinom}, the probability distributions at a primary neutron energy of \SI{600}{\MeV} are shown.
The double plane efficiency factor is proportional to the neutron reaction cross section and thus energy-dependent, ranging from \SI{10}{\percent} to \SI{12.2}{\percent}.
At \SI{600}{\MeV}, $p_{DP}$ is \SI{11.03(1)}{\percent}.

While the binomial behavior is expected, it is important to be explicitly aware of the implications for multi-neutron events:
As expected by the Lambert-Beer law for the attenuation of uncharged particles in media, increasing the detector depth quickly leads to diminishing returns for the detection of a single neutron.
However, for multi-neutron detection, this marginal change enters a power law.
If, for example, only seven double planes are used, four out of four neutrons will react only in \SI{9}{\percent} of all cases.
This can be quintupled to \SI{46}{\percent} by doubling the number of double planes.
In addition, a larger detector depth also results in fewer events where not all neutrons have reacted, which eases event reconstruction.

These basic considerations of the reaction probability show that experiments targeting the prestigious detection of four-neutron events will be quite challenging with less than half of the full detector depth --- without even looking at reconstruction efficiency.
From this standpoint, it seems well justified to target 30 double planes for the final detector depth.

\begin{figure}[htb]
\includegraphics[width=\columnwidth]{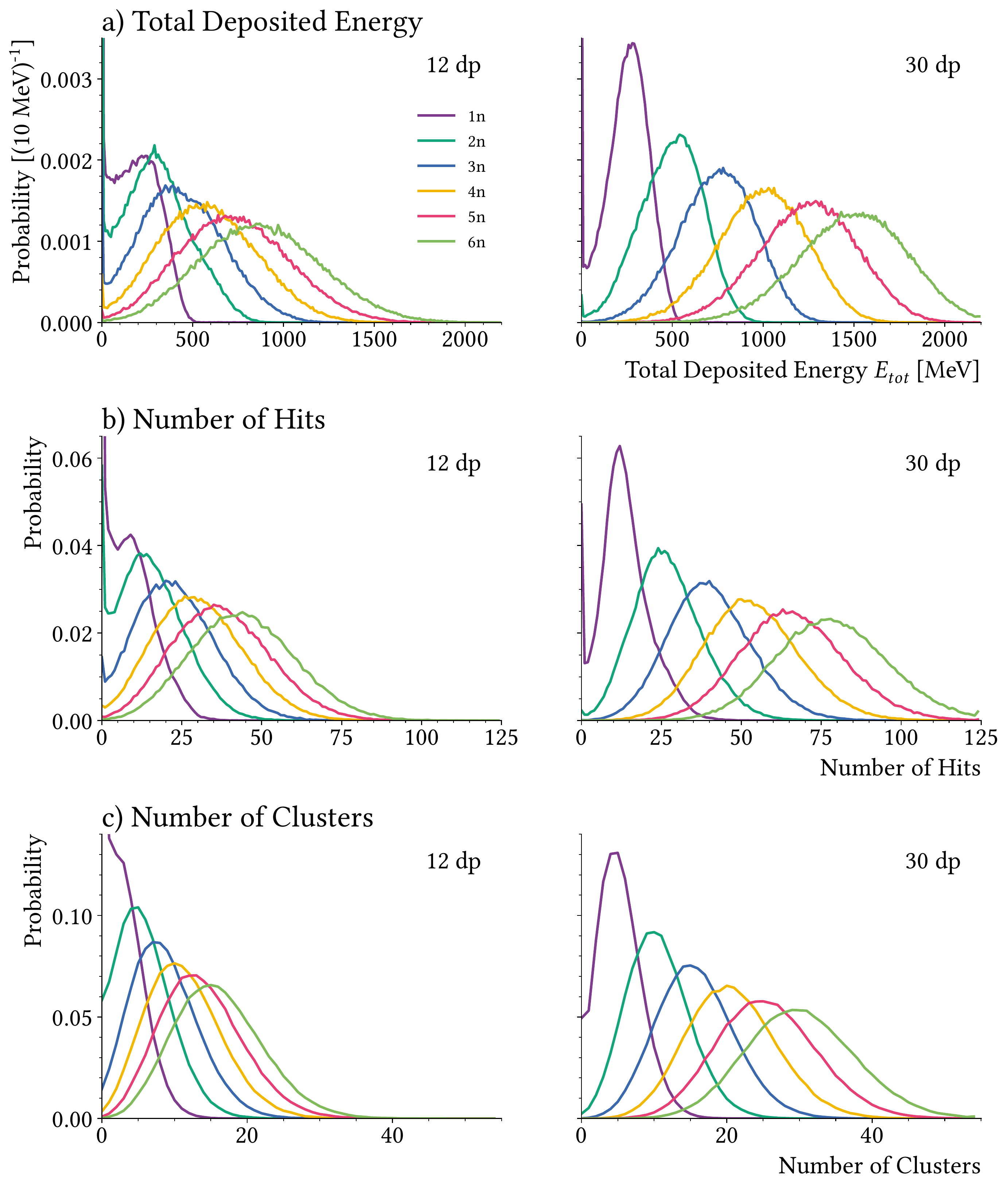}
\caption{Probabilities for total deposited energy, number of hits, and the number of clusters for different numbers of incoming neutrons in 12 and 30 double planes for a neutron energy of \SI{600}{\MeV}.
Note that while the average deposited energy and the number of clusters are correlated to the number of incoming neutrons, the resolution is not high enough to separate the individual channels event-by-event with cuts.
The reduced separation for fewer double planes is a convolution of a reduced probability for all neutrons to react and a lower chance for complete energy deposition.\label{f:eventfeatures}}
\end{figure}

\section{Multiplicity}\label{s:multiplicity}
Determining the multiplicity event-by-event is a classification problem.

We have investigated several classification methods for different scenarios.
The original design document included a method based on separating multiplicities by imposing hard cuts in a 2D histogram of deposited energy and number of clusters.
We found that the correlation between these quantities is not always sufficient for a clean separation.
An approach inspired by Bayesian statistics can give probabilities for each event with little effort by treating the individual 1D histograms of the number of hits, the number of clusters, and the deposited energy as likelihoods.
Machine learning with predefined algorithms from the scikit-learn library \cite{scikit-learn} and neural networks with Keras \cite{Chollet2015} were also investigated.

Before models can be trained, it is crucial to choose or \emph{engineer} informative, discriminating, and independent features.
For the multiplicity reconstruction, we constructed several different sets of features.
These sets have varying numbers of attributes and dimensions, indicated in parentheses after their designation.
This is often referred to as the \emph{shape} of the input in machine learning terms.
For example, $(n_{DP} \cdot 100, 2)$ describes a shape with 6000 values (for $n_{DP} = 30$ double planes) arranged in two dimensions, also called \emph{matrix} or \emph{rank-2} tensor.
\begin{description}\label{s:features}
\item[Trifeature: $(3)$]{From the many individual hits in the detector, the number of hits, the total deposited energy, and the number of clusters are calculated.  The distributions broaden for higher neutron multiplicities, as they are convolutions of the one-neutron distribution, see \cref{f:eventfeatures}. All detailed information about patterns in the detector and all-time information is discarded.}
\item[Bars: $(n_{DP} \cdot 100, 2)$]{For each of the up to 3000 scintillator bars, time and energy are recorded. This format retains some position information, as the position in the array corresponds to a specific bar, which is always at a specific position. Note that most entries will be zero for any given event.}
\item[Bars+Tri: $(n_{DP} \cdot 100 \cdot 2 + 3)$]{Like the Bars dataset, but with the Trifeature set added to help the model along.}
\item[Pixels: $(50,50,n_{DP} \cdot 2,2)$]{As the bars have a square profile of \SI{5}{cm} and the position resolution within the bar is even better, one can interpret the detector as a 3D image of 50 x 50 x 60 pixels with the two \enquote{color} channels time and energy. This format is inefficient and requires a significant amount of storage without compression, as the overwhelming number of pixels will be zero.}
\end{description}

Models were trained and evaluated on a train-test-split of data simulated as described in \cref{s:degen} for one to four neutrons.
An example of the resulting confusion matrices is given in \cref{t:confmat}.
To condense the performance down to a single number, all models were evaluated with the balanced accuracy score (BAC) from scikit-learn, which is the average recall obtained for each class with possible imbalances of the dataset taken into account.
A perfect reconstruction would score 100\%.
Balanced accuracy scores and training time for a selection of models are given in \cref{t:multresults}.

\subsection{Overall multiplicity}\label{s:expmult}
\begin{figure}
\centering
\includegraphics[width=\columnwidth]{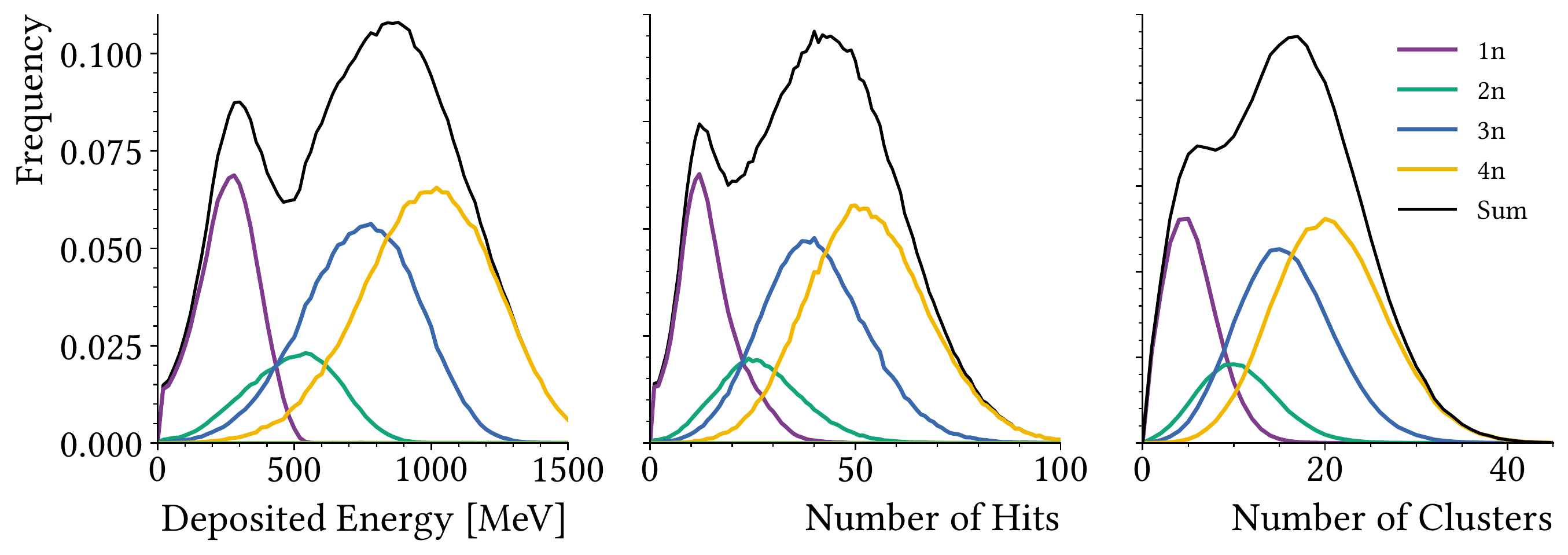}
\caption{Simulated histograms for an experiment with 20\% 1n, 10\% 2n, 30\% 3n, and 40\% 4n incoming neutrons (solid black). From these cumulative distributions, the constituting parts (solid colored) can be extracted.
\label{f:expdist}}
\end{figure}

In an experiment, the measured distributions $P_{exp}$ for total deposited energy, the number of hits, and the number of clusters are accumulated from many events with varying numbers of incoming neutrons.
Thus, the experimental distributions can be expressed as a sum of the individual distributions $P_n$ shown in \cref{f:eventfeatures} with experiment-specific weighting factors $a_n$.
\begin{equation}
P_{calc} = \sum_{n=1}^{n_{max}} a_n P_n
\end{equation}
These weighting parameters can be determined by minimizing the squared differences between the measured and simulated distributions, e.g.,
\begin{equation}
\begin{split}
f_{min}(a_1, ..., a_{n_{max}}) =
\sum \left( P_{calc}^{E_{dep}}  - P_{meas}^{E_{dep}}  \right)^2 \\ +
\sum \left( P_{calc}^{Hits}     - P_{meas}^{Hits}     \right)^2 +
\sum \left( P_{calc}^{Clusters} - P_{meas}^{Clusters} \right)^2
\end{split}
\end{equation}
An example is shown in \cref{f:expdist}.
The weighting parameters reflect the number of events with a specific multiplicity, which is one of NeuLANDs key deliverables.
They can also enter event-by-event multiplicity reconstruction as \emph{prior}, see \cref{s:bayes}.
This reflects that in an experiment with, e.g., 10\% 3n- and 90\% 4n events, a single event is much more likely to stem from four incoming neutrons.

\subsection{Calorimetric Method}\label{s:calorimetric}
\begin{figure}
\includegraphics[height=0.375\columnwidth]{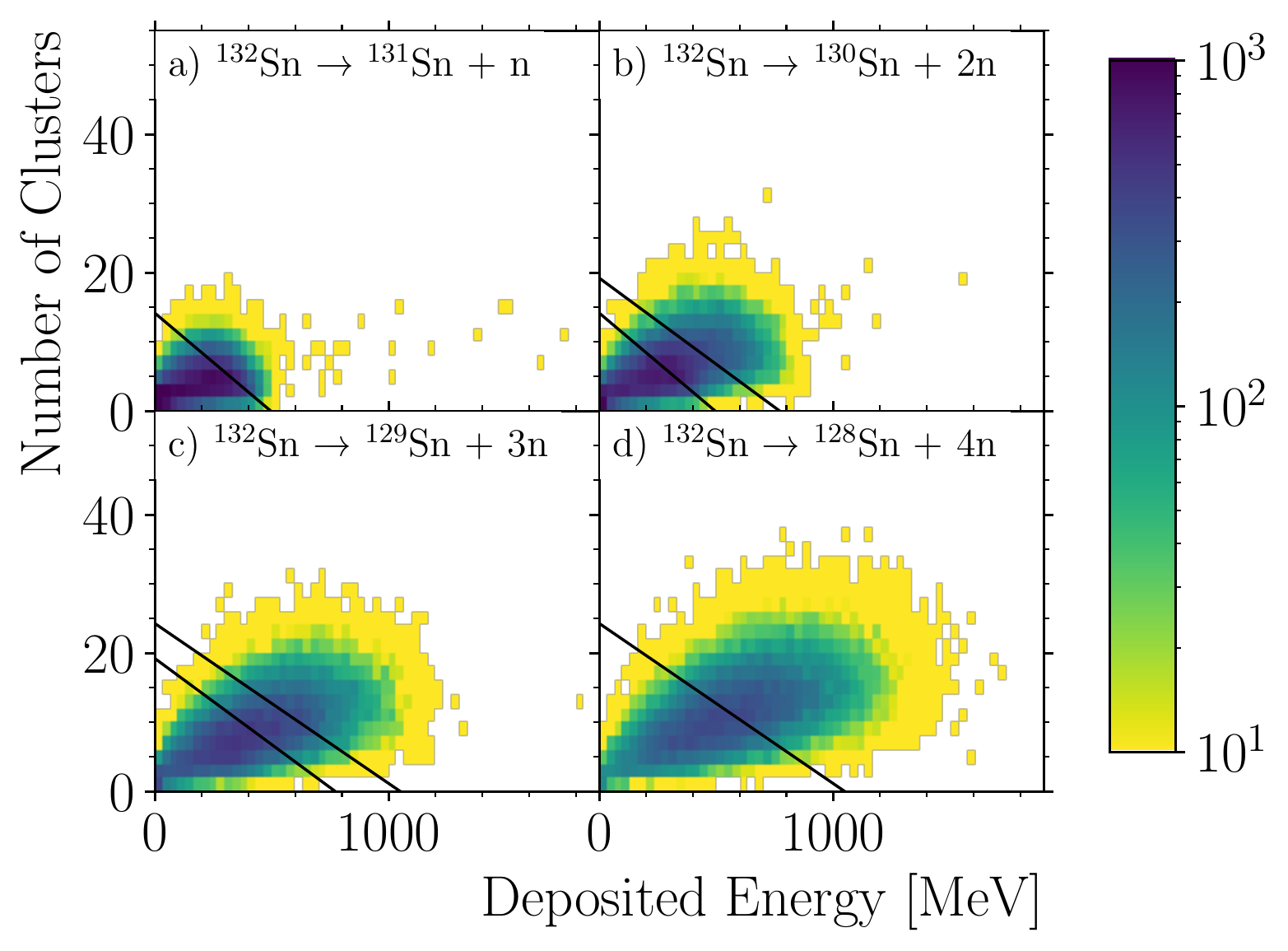}
\includegraphics[height=0.375\columnwidth]{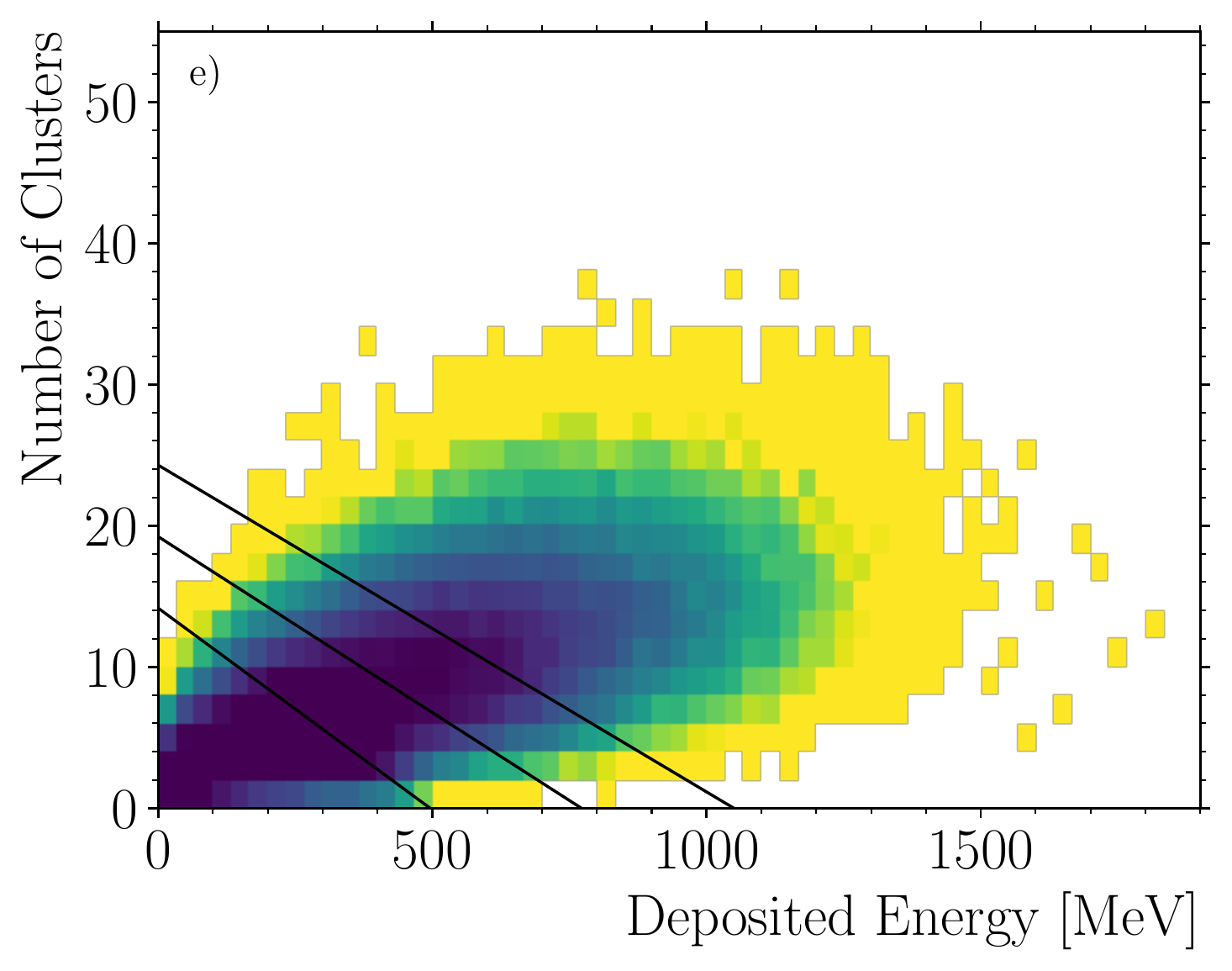}
\caption{Number of clusters versus the total deposited energy ($E_{dep}$) for \SI{600}{MeV} neutrons on 12 double planes. Quadrants (a) to (d) show the individual simulated data of neutron multiplicities between 1 and 4. In (e) the data of all neutron multiplicities has been combined. The lines represent the conditions which are applied to distinguish the different multiplicities, and are placed in a way to minimize false classifications, see \cref{s:calorimetric}.\label{f:calibr}}
\end{figure}

The original method for event-by-event multiplicity reconstruction is based on setting cuts in 2D-histograms where the number of clusters is plotted against the total deposited energy \cite{NeuLANDpaper, NeulandTDR}.
For each multiplicity, a designated zone in the histogram is created with cuts, see \cref{f:calibr}.
All events are classified by the multiplicity of the zone they fall into.
The placement of these cuts is optimized by minimizing the number of misclassified events.
This process can roughly be described as a support vector regression (or SVM) with special restrictions.
It is directly implemented into R3BRoot and uses ROOT files.
We include it in the list of processes using the Tri-Dataset, although it uses only two of the three values.

We found that this calorimetric method achieves acceptable balanced accuracy scores for the test case presented here and holds up surprisingly well compared to other models, see \cref{t:multresults}.
A problem with this solution is the minimization procedure: It requires settings, i.e., start parameters and limits for the parameters which match the beam energy, and the minimization can fail.
For example, due to the placement of the cuts at low energies, it does not handle two-neutron separation well.

The main limitation is the large overlap of the peaks in the histograms.
This is especially problematic if only 15 double planes are in place because then the accuracy drops due to the reduced calorimetric properties of the detector.
One can also further criticize the sharp transitions between the zones, as an event with slightly varying energy, e.g., in the range of the energy resolution, might suddenly be classified with a different multiplicity.

\subsection{Bayesian Statistics}\label{s:bayes}
A simple probabilistic approach can be implemented with Bayesian statistics.
Here, probabilities $P$ for hypotheses $H$ are calculated under the effect of data $\mathbf{E} = \{E_1, ... , E_k\}$
\begin{equation}
P(H|\mathbf{E}) = P(H) \frac{P(\mathbf{E}|H)}{\sum_{h} P(\mathbf{E}|H_h) P(H_h)}
\end{equation}
with
\begin{equation}
P(\mathbf{E}|H) = \prod_{E_i\,\in\,\mathbf{E}} P(E_i|H),
\end{equation}
where $P(H)$ is the probability before the consideration of data, called \emph{prior}; $P(H|\mathbf{E})$ the probability of $H$ after $\mathbf{E}$, called \emph{posterior}; and $P(E|H)$ the probability of observing $E$ given $H$, also known as the \emph{likelihood}.
Effectively, likelihoods can be multiplied with each other which, after normalization, results in a probability for each possible outcome.

The multiplicity reconstruction problem can be translated to this Bayesian domain language.
Hypotheses are the desired neutron multiplicities:
\begin{equation}
\mathbf{H} = \{0n, 1n, 2n, 3n, 4n, 5n\}
\end{equation}
The prior can be chosen from an external source, if available, see \cref{s:expmult}.
For the results presented here, we use the same prior for all hypotheses.
The number of hits, the number of clusters, and the total deposited energy shown in \cref{f:eventfeatures} are the likelihoods entering the calculations.
For example, $P(E_\text{dep}=\SI{100}{\MeV}|1n)$, that is the likelihood to find a deposited energy of \SI{100}{\MeV} if one neutron has reacted, can be directly read off the histogram for the deposited energy for one neutron.
Note that in principle, the individual probabilities would have to be independent, which is not fulfilled perfectly here.

This method has several significant advantages:
Training and testing of the basic dataset require only minimal computational resources due to the highly optimized NumPy library \cite{harris2020array}.
In addition, there is no minimization step that could fail, and the method requires no settings or variable tuning.
The whole process can also be easily converted and included in the \texttt{C++} analysis software, which enables high throughput when analyzing millions of events.
As it returns probabilities and can directly include priors, it is also well suited for usage with event-wide multiplicity distributions, see \cref{s:expmult}.
Finally, it only uses the unscaled Tri dataset, so it is robust against changes to the hit pattern due to, e.g., changes in the neutron cone opening angle, and less susceptible against simulation-related problems like imperfect physics models.

This method achieves a good, but not the best 30dp-4n-PH-BAC performance of \SI{79}{\percent} for the \SI{600}{MeV} scenario tested here.

\begin{table*}
\caption{Confusion matrices (neutron separation matrices) for up to three, four, and five primary hits ($N_{PH}$).
Rows display the actual number of primary hits (the \emph{true} multiplicity) and columns the number of primary hits derived with the Bayes algorithm (the \emph{predicted} multiplicity).
The correct assignments are highlighted in bold and the balanced accuracy score is given below the method name.
Values are given in percent.
In addition, the number of events used to evaluate the performance of the trained model is given.
Notice that the multiplicity range the algorithm is tasked to reconstruct has a significant impact.
If, for example, a multiplicity of five or higher can be excluded due to external constraints like reaction mechanics or information from other detectors, a multiplicity of four can be reconstructed more accurately.
Neutrons were simulated with 600\,MeV and a relative energy of 500\,keV with a \isotope[132-n]{Sn} fragment.
NeuLAND with 30 double planes was located at a distance of 15\,m to the target.
The distance between target and NeuLAND was filled with air and a 4\,mm steel window.
Simulated with \textsc{Geant4} using the \texttt{QGSP\_INCLXX\_HP} physics list.
\label{t:confmat}}
\begin{singlespace}
\centering
\resizebox{0.8\width}{!}{
\begin{tabular}{cc|S[table-format=2]S[table-format=2]S[table-format=2]S[table-format=2]|S[table-format=2]}
\toprule
\multicolumn{2}{c|}{Bayes} & \multicolumn{4}{c|}{predicted} & {Events} \\
\multicolumn{2}{c|}{85} & 0 & 1 & 2 & 3 & {[$10^3$]}\\
\midrule
\multirow{4}{*}{\rotatebox[origin=c]{90}{true}}
 & 0 & \textbf{100} &    &    &    &  2 \\
 & 1 &     & \textbf{91} & 9  &    & 42 \\
 & 2 &     & 14 & \textbf{69} & 17 & 42 \\
 & 3 &     &  1 & 19 & \textbf{80} & 34 \\
\bottomrule
\end{tabular}
}
\resizebox{0.8\width}{!}{
\begin{tabular}{cc|S[table-format=2]S[table-format=2]S[table-format=2]S[table-format=2]S[table-format=2]|S[table-format=2]}
\toprule
\multicolumn{2}{c|}{Bayes} & \multicolumn{5}{c|}{predicted} & {Events} \\
\multicolumn{2}{c|}{79} & 0 & 1 & 2 & 3 & 4 & {[$10^3$]}\\
\midrule
\multirow{5}{*}{\rotatebox[origin=c]{90}{true}}
 & 0 & \textbf{100} &    &    &    &    &  2 \\
 & 1 &     & \textbf{91} &  9 &    &    & 42 \\
 & 2 &     & 14 & \textbf{69} & 17 &    & 42 \\
 & 3 &     &  1 & 19 & \textbf{59} & 22 & 41 \\
 & 4 &     &    &  2 & 22 & \textbf{76} & 33 \\
\bottomrule
\end{tabular}
}
\resizebox{0.8\width}{!}{
\begin{tabular}{cc|S[table-format=2]S[table-format=2]S[table-format=2]S[table-format=2]S[table-format=2]S[table-format=2]|S[table-format=2]}
\toprule
\multicolumn{2}{c|}{Bayes} & \multicolumn{6}{c|}{predicted} & {Events} \\
\multicolumn{2}{c|}{74} & 0 & 1 & 2 & 3 & 4 & 5 & {[$10^3$]}\\
\midrule
\multirow{6}{*}{\rotatebox[origin=c]{90}{true}}
 & 0 & \textbf{100} &    &    &    &    &    &  2 \\
 & 1 &     & \textbf{91} &  9 &    &    &    & 42 \\
 & 2 &     & 14 & \textbf{69} & 17 &    &    & 42 \\
 & 3 &     &  1 & 19 & \textbf{59} & 21 &    & 42 \\
 & 4 &     &    &  2 & 22 & \textbf{51} & 25 & 40 \\
 & 5 &     &    &    &  3 & 23 & \textbf{73} & 31 \\
\bottomrule
\end{tabular}
}
\end{singlespace}
\end{table*}

\subsection{Simple Machine Learning with scikit-learn}\label{s:multsl}
Scikit-learn \cite{scikit-learn} is a widely used Python machine learning library that includes a wide range of classification, regression, clustering, and other models.
Version \texttt{0.21.3} used here includes 32 applicable classifier models, 10 feature scalers including unscaled input, and 6 multi-model-classifiers.

Using the Tri, Bars, and Bars+Tri dataset as input, we tested all possible dataset-scaler-model combinations with default settings.
Some models work better with or require scaled inputs.
Others do not scale well with the number of events, for some, the training time also depends on the scaler used.
As the training time can vary by orders of magnitude, we applied a strict time limit to cancel and retry training the specific combination with a tenth of the data.

To our surprise, we found that many models perform roughly the same, at a level that is only slightly above Calorimetric or Bayes, while requiring significantly more training time.
A selection is given in \cref{t:multresults}.
All models performed worse when working with the thousands of inputs of the Bars dataset than with the three values of the Tri dataset.
Usage of the combined Bars+Tri dataset did not lead to better performance but required drastically more processing time.
We also found that some models, especially the Quadratic Discriminant Analysis, perform significantly worse when trained on the number of primary neutrons instead of the (actually detected) number of primary hits.

\subsection{Neural Networks with Keras}\label{s:mulker}
The Keras framework \cite{Chollet2015} was used to build simple neural networks.

Configurations with dense Relu layers, Softmax (SM) activation, Adam optimizer, and the Categorical Crossentropy loss function were used, see \cite{Chollet2015}.
At first, we tackled the hyperparameter optimization with Keras-tuner, optimizing layer sizes of thousands of Relu nodes for the Max-Abs scaled Bars and Bars+Tri datasets.
However, we quickly found that in our case these layer sizes lead to drastic overfitting with training accuracies of \SI{99}{\percent} but decreasing validation accuracy.
Instead, configurations of two layers with 100 and 10 nodes (R100, R10, SM, Adam), as well as a single 10 node layer (R10, SM, Adam) performed as well as the best scikit-learn models.

The single most surprising result was that, without any Relu-layer at all, just the Softmax-Layer with Adam optimizer (SM, Adam) achieved a great balanced accuracy score of \SI{83}{\percent} for 4n on 30dp when reconstructing PH on the unscaled Tri dataset.
It is much faster to train and run than all scikit-learn models.
This performance is due to the highly optimized TensorFlow code \cite{tensorflow2015-whitepaper}.
However, for best data-evaluation performance it might be necessary to integrate the trained network into R3BRoot without its Python API to avoid the slow (ROOT-) data preprocessing in Python.

\subsection{Summary}
Evaluating the performance of the described model-scaler combinations leads to some unexpected results, see \cref{t:multresults}.

First, a BAC of \SI{83}{\percent} seems to be the hard limit for the reconstruction of the number of primary hits for one to four primary neutrons with 30 double planes (30dp-4n-PH-BAC) and \SI{69}{\percent} for 12 double planes.
This value is reached by several models and might originate from the properties of the simulated interactions.

Second, the use of more features does not provide a guaranteed advantage. For example, using the 3000 time and 3000 energy features in the Bars dataset even leads to a significant under-performance for all models, highlighting the need for feature engineering (or better-designed models).
Also, the combination of the derived features (Tri) and the full detailed pattern (Bars) did not lead to outperformance, but just to an increase in processing time.
Due to this lack of performance, the Pixel-Dataset was not tested for pure multiplicity reconstruction.

The calorimetric method held up surprisingly well, almost matching the hard limit for deducing $N_{PH}$, only falling behind when predicting the number of primary neutrons for 12 double planes.
Its main drawback is the fitting algorithm, which has parameters and can fail.

The simple Bayesian model performs admirably well, its main advantages are its simplicity and robustness.
It also natively integrates well to using prior knowledge from the overall multiplicity distribution.
We nominate it as the default model.

As expected, we observed highly mixed behavior ranging from total failure to the hard limit for scikit-learn models.
Scalers do have an impact on, e.g., the SGDClassifier, where they drastically reduce training time; and the LinearDiscriminantAnalysis, where they drastically improve accuracy (not shown in \cref{t:multresults}).

Note that the multiplicity determination accuracy also depends on the maximum multiplicity.
As shown in \cref{f:eventfeatures}, the overlap between the basic observables number of hits, number of clusters, and total deposited energy is large.
If the experiments can only produce up to three neutrons, the option to reconstruct four and five neutrons can be eliminated, which will lead to much higher accuracy.

\begin{table*}
\caption{Balanced Accuracy Scores (BAC) for the reconstruction of multiplicities for 1 to 4 incoming neutrons.
Manually implemented, default Scikit-Learn, and Keras-based Models were systematically trained and evaluated with all Dataset-Scaler combinations on datasets with the same number of events for each multiplicity.
Both the number of incoming neutrons ($N_{PN}$) and the number of primary hits ($N_{PH}$) were separately used as label.
Scikit-Learns Models with low performance are omitted. \enquote{All} scalers include Unscaled (Unsc), standard scaling (Std), min-max scaling (MM), max-abs scaling (MA), robust scaling (Rob),  power transformation: Yeo-Johnson (YJ), power transformation: Box-Cox (BC), quantile transformation: gaussian pdf (QG), quantile transformation: uniform pdf (QU), sample-wise L2 normalizing (L2), see \cite{scikit-learn, scaler} for details. Not used were Power Transformer Yeo-Johnson and Box-Cox.
\emph{Intrinsic efficiency} results are obtained using the known number of primary hits, which leads to a perfect score when reconstructing itself, but as not all incoming neutrons will react in the detector, to a lower score when trying to reconstruct the number of incoming neutrons. See text for details.\label{t:multresults}}
\centering
\begin{singlespace}
\begin{tabular}{l|ll|S[table-format=2]S[table-format=2]|S[table-format=2]S[table-format=2]}
\toprule
Model                        & Input    & Scaler            & \multicolumn{4}{c}{BAC [\%]} \\
                             &          &                   & \multicolumn{2}{c}{12dp} & \multicolumn{2}{c}{30dp} \\
                             &          &                   & {$N_{PN}$} & {$N_{PH}$} & {$N_{PN}$} & {$N_{PH}$} \\
\midrule
Intrinsic efficiency         & -        & -                 & 44 &100 & 84 &100 \\

\midrule
Calorimetric                 & Tri      & Unsc              & 41 & 66 & 70 & 82 \\
Bayes                        & Tri      & Unsc              & 57 & 67 & 74 & 79 \\
\midrule
MLP Classifier               & Tri      & All               & 57 & 66 & 78 & 83 \\
(Hist) Gradient Boosting Class.     & Tri & All             & 57 & 66 & 78 & 83 \\
Logistic Regression CV       & Tri      & All               & 56 & 66 & 76 & 83 \\
Quadratic Discriminant Analysis & Tri & Std, Rob, QG, YJ    & 36 & 66 & 56 & 83 \\
Quadratic Discriminant Analysis & Tri & Unsc, MA, L2, MM, QU & 20 & 65 & 20 & 83 \\
Linear Discriminant Analysis & Tri      & QG, YJ            & 56 & 66 & 76 & 82 \\
\midrule
Keras (SM, Adam)             & Tri      & Unsc              & 56 & 67 & 76 & 83 \\
Keras (R10, SM, Adam)        & Tri      & Unsc              & 57 & 66 & 76 & 83 \\
Keras (R100, R10, SM, Adam)  & Tri      & Unsc              & 54 & 64 & 78 & 82 \\

\midrule
Bernoulli NB                 & Bars     & All               & 56 & 62 & 67 & 70 \\
MLP Classifier               & Bars     & SS, MA, MM        & 56 & 59 & 62 & 66 \\
Extra Trees Classifier       & Bars     & QU, QG, L2, Unsc  & 59 & 59 & 65 & 65 \\
Random Forest Classifier     & Bars     & QU, QG, Rob, Unsc & 60 & 58 & 63 & 65 \\
\midrule
Keras (SM, Adam)             & Bars     & MA                & 56 & 62 & 50 & 54 \\
Keras (R10, SM, Adam)        & Bars     & MA                & 58 & 63 & 71 & 74 \\
Keras (R100, R10, SM, Adam)  & Bars     & MA                & 59 & 66 & 71 & 75 \\
Keras (R3000, R50, SM, Adam) & Bars     & MA                & 56 & 63 & 69 & 75 \\

\midrule
Linear Discriminant Analysis & Bars+Tri & QG                & 59 & 69 & 75 & 82 \\
Bagging Classifier           & Bars+Tri & All w/o L2        & 56 & 65 & 76 & 81 \\
Extra Trees Classifier       & Bars+Tri & MA, MM            & 60 & 64 & 75 & 80 \\
Random Forest Classifier     & Bars+Tri & All w/o L2        & 60 & 63 & 75 & 79 \\
Nearest Centroid             & Bars+Tri & MA, MM            & 52 & 63 & 72 & 77 \\
MLP Classifier               & Bars+Tri & QG                & 60 & 69 & 69 & 75 \\
\midrule
Keras (SM, Adam)             & Bars+Tri & MA                & 59 & 67 & 76 & 82 \\
Keras (R10, SM, Adam)        & Bars+Tri & MA                & 60 & 69 & 77 & 83 \\
Keras (R100, R10, SM, Adam)  & Bars+Tri & MA                & 60 & 69 & 76 & 82 \\
Keras (R3000, R50, SM, Adam) & Bars+Tri & MA                & 57 & 66 & 72 & 80 \\
\bottomrule
\end{tabular}
\end{singlespace}
\end{table*}

\section{First interaction points}\label{s:fip}
The first interaction points are required to determine the four-vectors of the primary neutrons.
There are several possible approaches to this problem.
In an ideal case, one would simply use the full Bar or Pixel shaped data as input (see \cref{s:multiplicity}), similar to a 3D image, and the model would point out the exact positions, likely even with higher precision as the intrinsic resolution of the scintillator would normally allow for.

Here, we present a different approach.
All detected interactions are already represented by clusters, and the task is now reduced to classify them into primary and secondary.
As input for this binary classification, either a 2D or 3D representation of the individual hits, i.e., a subset of the full detectors, or derived features representing meaningful attributes of the cluster can be used.

\subsection{Cluster classification}\label{s:clusterfeatures}
\begin{figure*}
\centering
\begin{tikzpicture}[x=\textwidth,y=-\textwidth]
\node[anchor=north west] at (0,0) {\includegraphics[width=\textwidth]{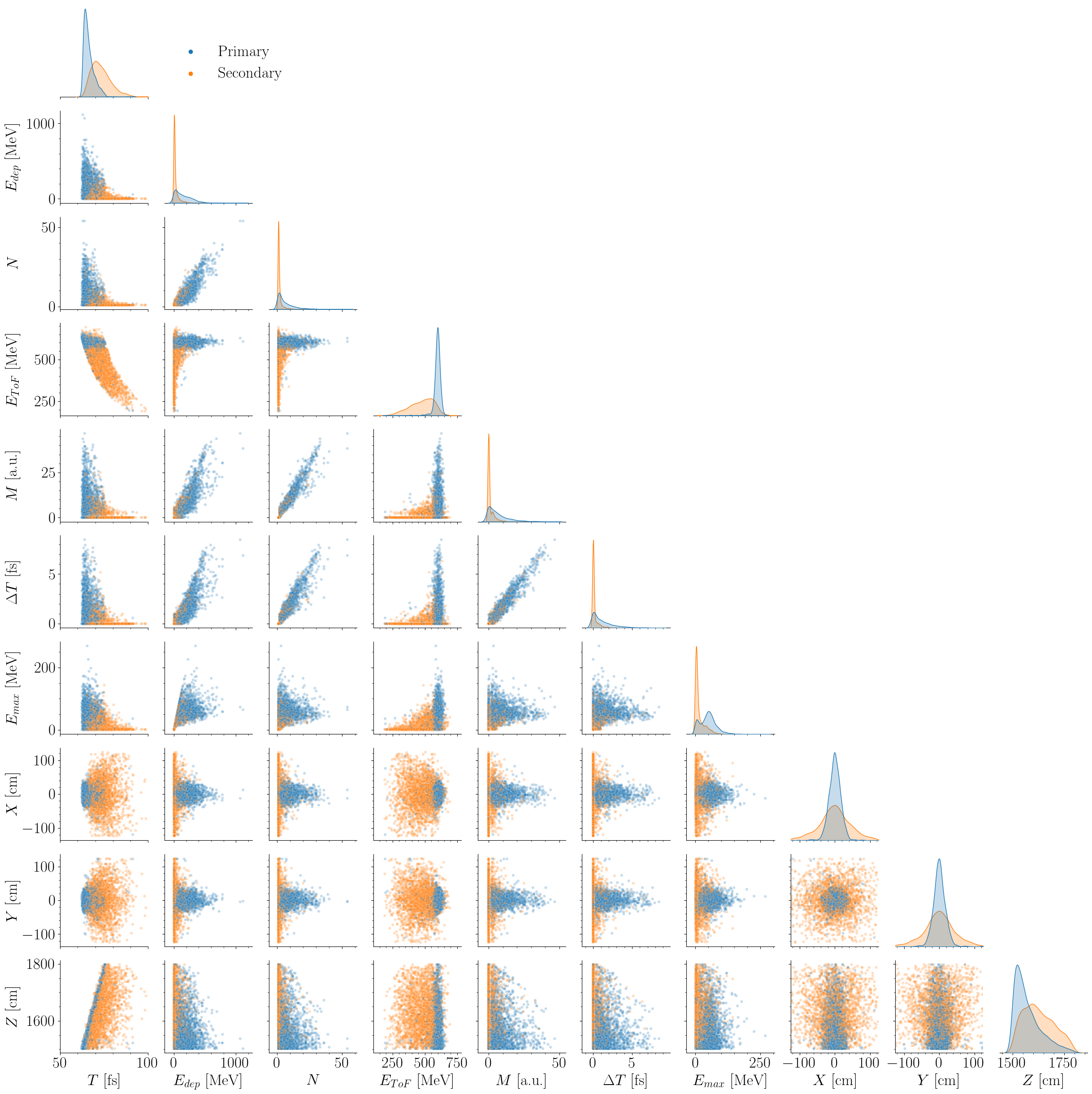}};
\node[anchor=north east] at (\textwidth,0) {
	\begin{tikzpicture}[x=\columnwidth,y=-\columnwidth]
	\node [anchor=north west] at (0,0) {\includegraphics[width=0.49\columnwidth]{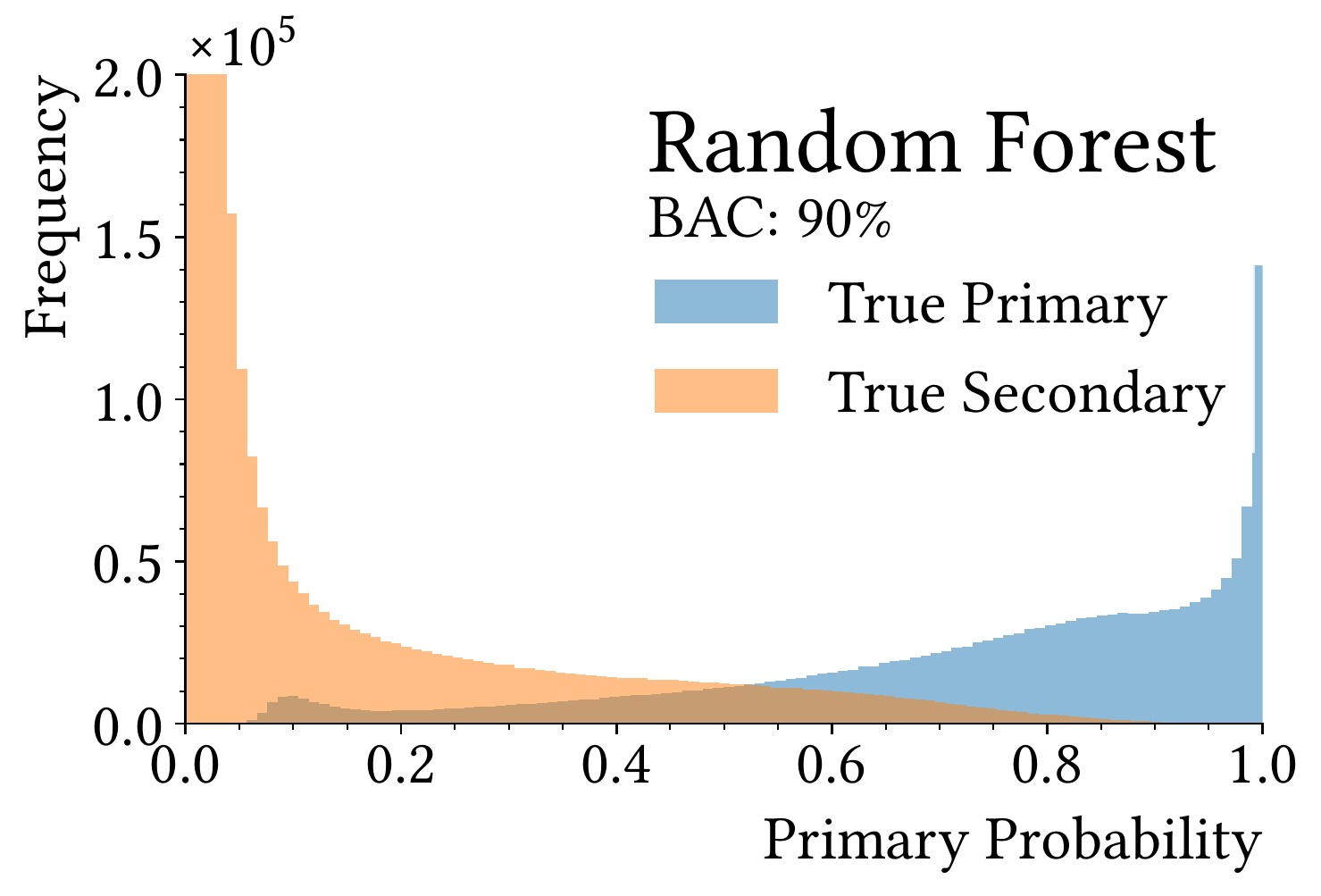}};
	\node [anchor=north east] at (\columnwidth,-0.00255) {\includegraphics[width=0.49\columnwidth]{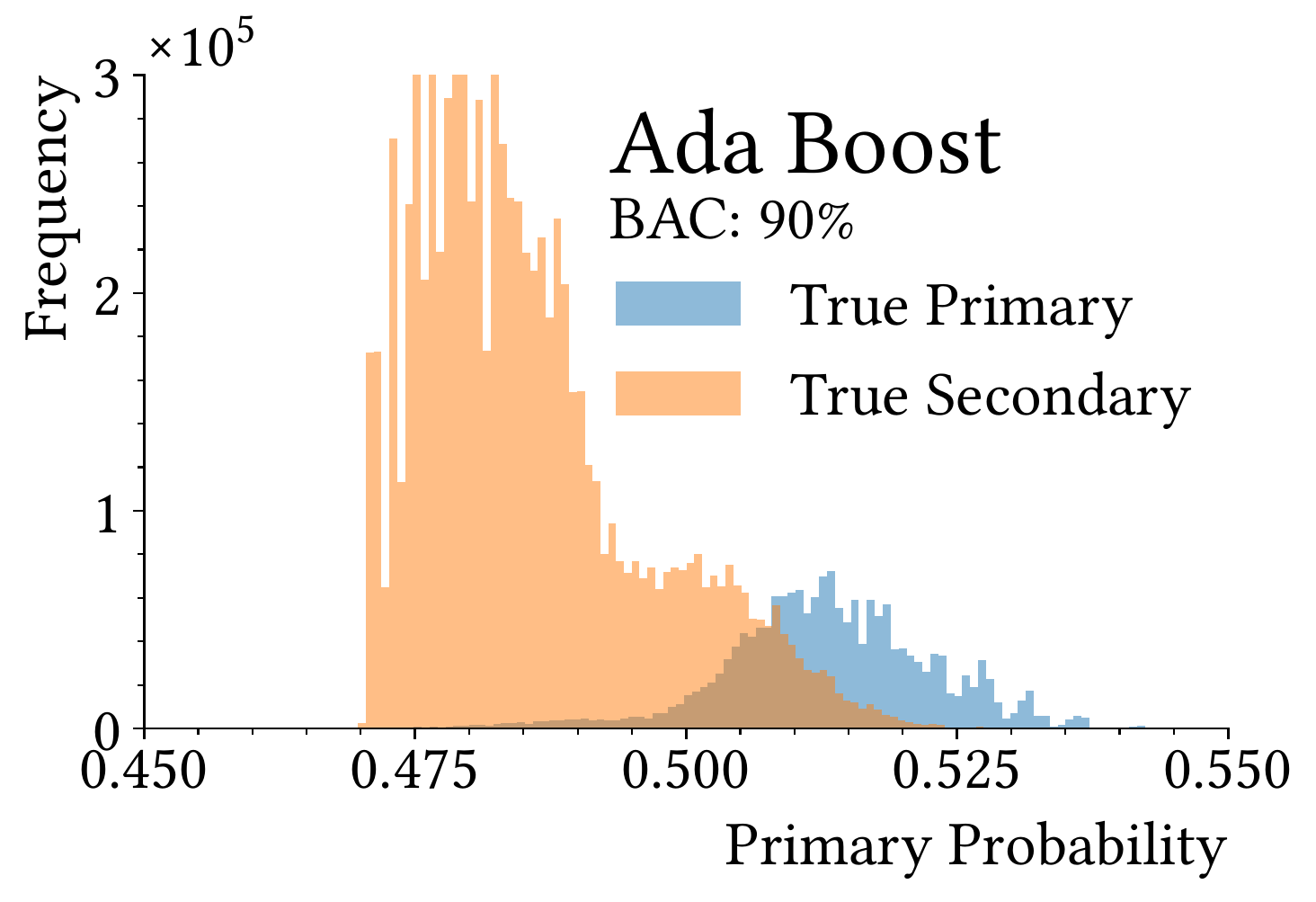}};
	\node [anchor=north west] at (0,0.33) {\includegraphics[width=0.49\columnwidth]{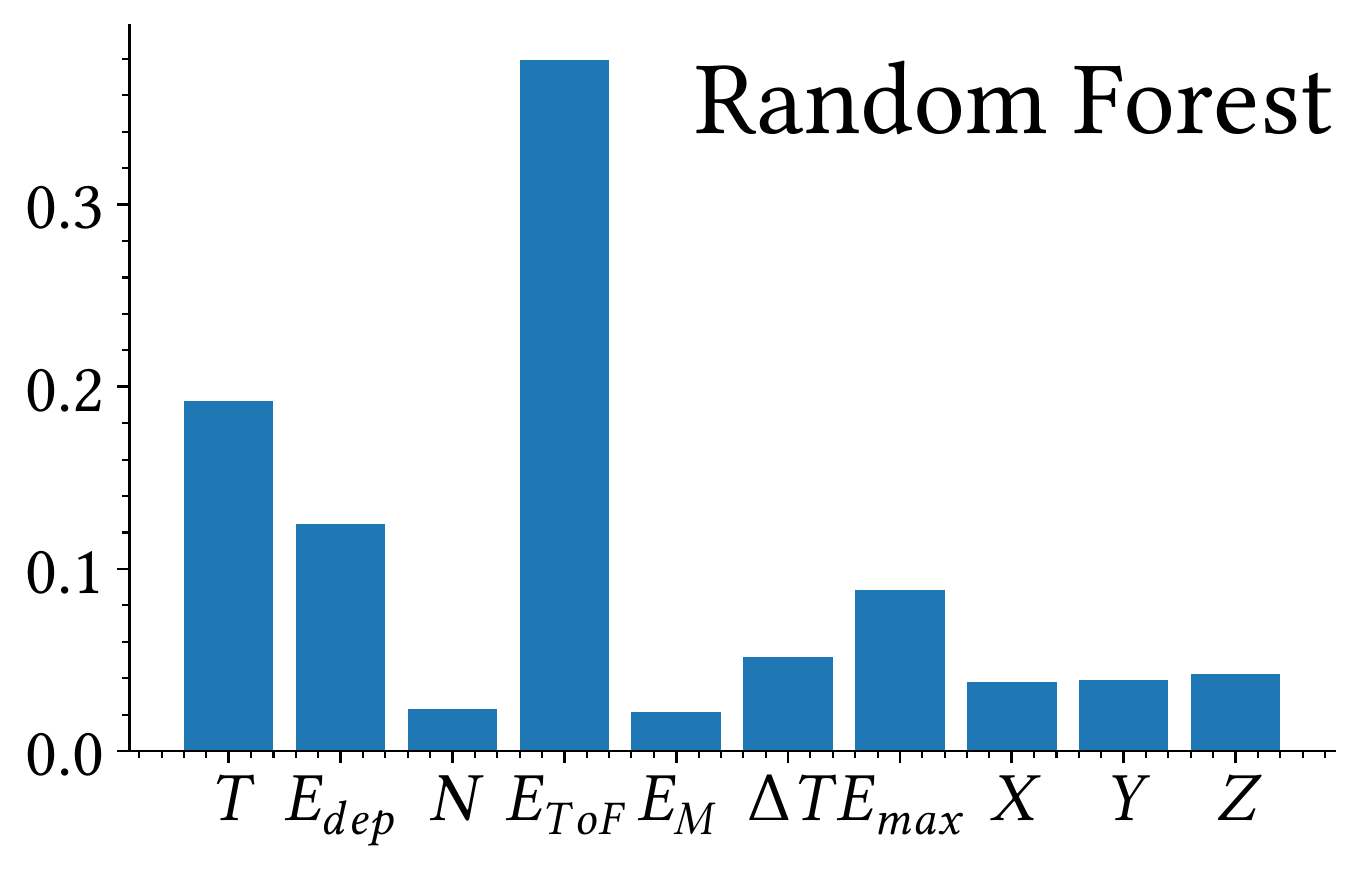}};
	\node [anchor=north east] at (\columnwidth,0.33) {\includegraphics[width=0.49\columnwidth]{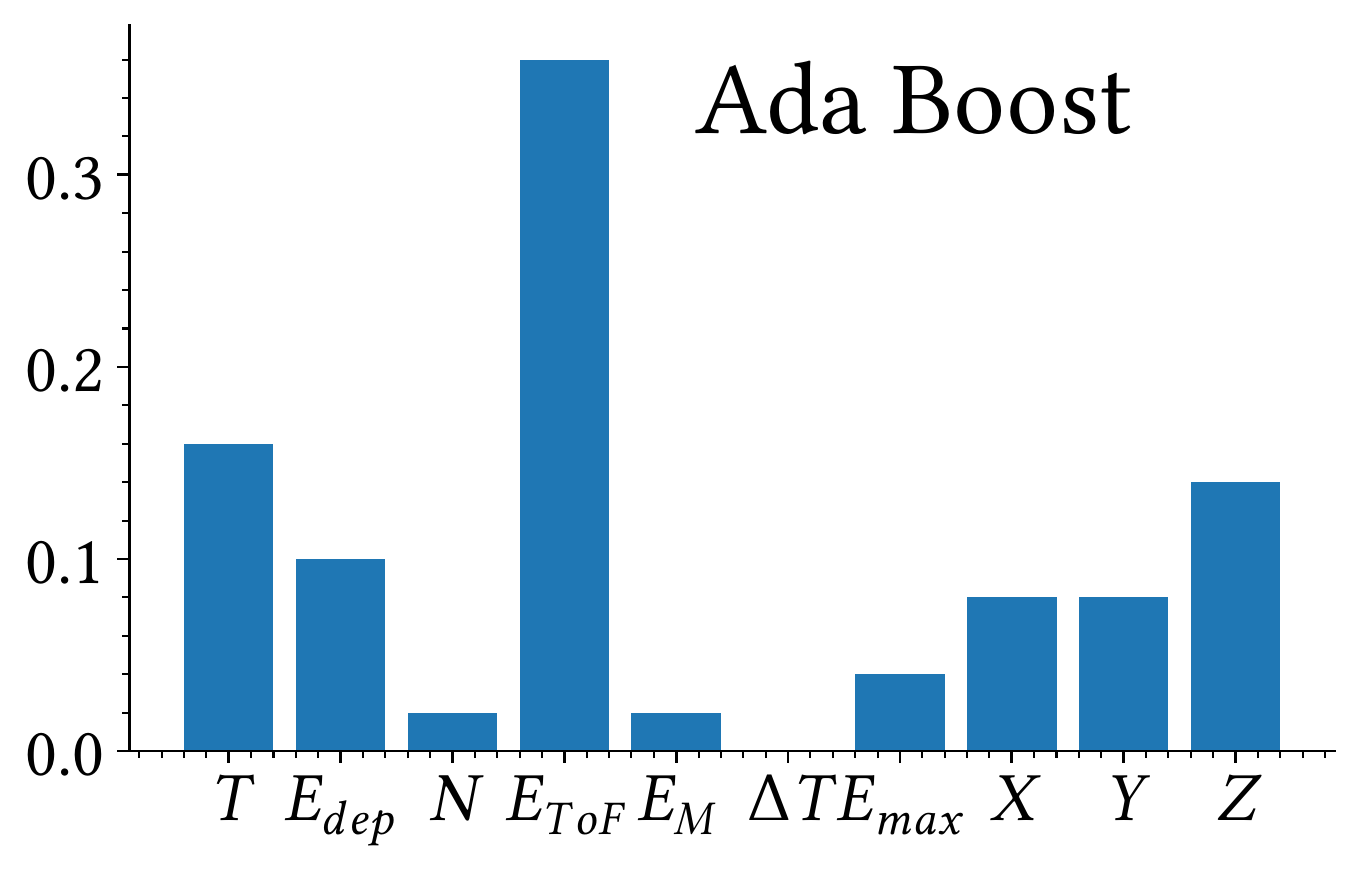}};
	\node [anchor=north west, text width=\columnwidth, align=justify] at (0,0.62) {\caption{Probability distributions (top) and importance of features (bottom) for the Random Forest (left) and Ada Boost classifier (right) from the scikit-learn library. Note that the impact of the features is quite similar for both classifiers. The energy from time-of-flight $E_{ToF}$ is by far the most important.}\label{f:clusterreco}};
	\end{tikzpicture}
};
\end{tikzpicture}
\caption{
Relationships between features of primary (blue) and secondary (orange) clusters. Note that, for example, the $X$-$Y$-distribution of primary clusters reflects the opening angle of the incoming neutrons. The $Z$-distribution reflects the exponential decrease in intensity (Lambert-Beer law). In the $T$-$Z$ plot, the longer flight time of neutrons that react in the back of the detector can be seen. The energy from time of flight $E_{ToF}$ removes this dependency and is a good indicator.
\label{f:clusterfeatures}}
\end{figure*}

Some properties of clusters might be good indicators for classification between primary and secondary clusters.
A pair plot of these features is shown in \cref{f:clusterfeatures}.
The details are discussed in the following:

\begin{description}
\item[Time-of-Flight $T$]{
    The Time-of-Flight $T$ is the time from the start signal (emission of the neutron) to the detected time of the first hit in the cluster.
    In many reactions, neutrons are emitted with a narrow kinetic energy window and thus arrive at the scintillators in a narrow time window.
    It is unlikely that clusters with a larger ToF stem from these primary neutrons.
    Note that the acceptable ToF depends on the flight distance as can be seen in the $T$-$Z$ pair plot.
    This is corrected for in $E_{ToF}$.}
\item[Deposited Energy $E_{dep}$]{
    The cluster energy $E_{dep}$ is the summed energy of the individual hits ($E_{dep} = \sum_{h} E_h$).
    There is an overabundance of secondary clusters with energies below \SI{10}{\MeV}. Clusters with an energy over \SI{150}{\MeV} are likely primary, see \cref{f:clusterfeatures}.}
\item[Size $N$]{
    The cluster size $N$ denotes the number of hits grouped in one cluster.
    Smaller clusters are more likely to be secondary, with an over-proportional number of both primary and secondary one-hit clusters.
    The likelihood of larger secondary clusters drops faster than for larger primary clusters, with a crossing at around size $N=7$.}
\item[Energy from Time-of-Flight $E_{ToF}$]{
    Assuming the cluster is created by a neutron stemming from the target, the time-of-flight, and the position of the first hit in the cluster can be used to calculate the neutron kinetic energy: $E_{ToF} = (\gamma - 1) m_n c^2$ with $\gamma=(1-\frac{X^2 + Y^2 + Z^2}{T^2 c^2})^{-\frac{1}{2}}$ and where $m_n$ is the neutron mass.
    In the physics cases simulated here, neutrons are emitted within a narrow energy window (Full Width at Zero Intensity < \SI{10}{\percent}), thus primary clusters must have a corresponding $E_{ToF}$.
    The energy from time of flight is connected to the time-of-flight $T$, however secondary clusters that would fall within the time-of-flight acceptance interval are rarely located at positions that match the required $E_{ToF}$.
    This makes the energy from time of flight a better feature for further analysis and even a candidate for a simple cut.}
\item[Cluster Energy Moment $M$]{
	Energy depositions in the cluster are not distributed evenly. A proton, for example, might deposit most of its kinetic energy at the end of its flight path (Bragg peak). The cluster energy moment $M$ can be defined as
	\begin{equation}
	M = \frac{ \sum_{h} \left| \mathbf{x}_h - \mathbf{x}_{EC} \right| E_h }{ E_{dep}} \text{\quad with \quad} \mathbf{x}_{EC} = \frac{ \sum_{h} \mathbf{x}_h E_h}{E_{dep}},
	\end{equation}
	where $h$ are the individual hits with energy $E_h$ and position $\mathbf{x_h}$. Assuming the cluster is a Bragg track, $M$ is a measure for the initial energy of the proton.
	}
\item[Timespan $\Delta T$]{
    The cluster timespan, which is the time difference between the last and the first hit in the cluster, behaves like the cluster size.
    This is expected, as larger clusters are mostly created by protons passing through the planes.}
\item[Maximum Hit Energy $E_{max}$]{
	The largest individual energy deposition in the cluster.}
\item[Position $X, Y, Z$]{
	The $X$, $Y$, and $Z$ position of the first hit in the cluster. For horizontally oriented scintillator bars, $Y$ is given by the scintillator position and $X$ is calculated from the time difference of the PMT signals. Vise versa for vertically oriented bars. The $Z$ position is always given by the scintillator position and shows the expected logarithmic decline.}
\item[R-Value]{
    The R-Value, defined as $R = \frac{\left|{E_{ToF} - E_{Beam}}\right|}{E_{dep}}$, is a component of the reconstruction method originally proposed in the technical design report \cite{NeuLANDpaper, NeulandTDR}.
    For the calculation of $R$, external knowledge of the beam energy $E_{Beam}$ is required.
    }
\end{description}

A similar broad search for the best classification process as in \cref{s:multiplicity} has been performed.
Recent developments in the field of machine learning include fully automating the process of finding the best
model with the best settings.
Here, we also utilize \emph{Auto-Sklearn} \cite{NIPS2015_5872} and compare it to our manual search.
In the first step, few far-away outliers were excluded.
The reaction of primary neutrons can create several secondary particles, which can react in different parts of the detector.
This results in up to four times more secondary than primary clusters.
A random subset of secondary clusters was used to normalize the dataset, which was then split into train and test sets.
The features have then been used as input with different scaler-model combinations, results are given in \cref{t:clusresults}.
Again, there seems to be a hard limit just above 91\% accuracy with little variation between the best scikit-learn models and Keras-based neural networks of different sizes.

\begin{table}
\caption{Balanced Accuracy Scores (BAC) for the reconstruction of primary clusters for 1 to 4 incoming neutrons.
Before training and testing, the dataset was normalized to include the same number of primary and secondary clusters.
Default Scikit-Learn were systematically trained and evaluated with all Dataset-Scaler combinations, Keras was tested with the Robust Scaler.
Only the best results are shown. \enquote{All} scalers include Unscaled (Unsc), standard scaling (Std), min-max scaling (MM), max-abs scaling (MA), robust scaling (Rob), quantile transformation: gaussian pdf (QG), quantile transformation: uniform pdf (QU), sample-wise L2 normalizing (L2). Not used were Power Transformer Yeo-Johnson and Box-Cox.
See text for details.\label{t:clusresults}}
\centering
\resizebox{\columnwidth}{!}{
\begin{tabular}{llS[table-format=2.1]S[table-format=2.1]}
\toprule
Model                        & Scaler           & \multicolumn{2}{c}{4n BAC [\%]} \\
                             &                  & {12dp} & {30dp} \\
\midrule
Random Forest Classifier     & All              & 90.2 & 91.2 \\
Extra Trees Classifier       & All              & 90.1 & 91.1 \\
MLP Classifier               & All w/o L2, Unsc & 90.0 & 90.9 \\
SVC                          & QU, Std, Rob     & 89.6 & 90.6 \\
Ada Boost Classifier         & All              & 89.3 & 90.2 \\
\midrule
Auto-Sklearn                 & -                & 90.5 & 91.4 \\
\midrule
Keras (SM, Adam)             & Rob              & 88.2 & 88.3 \\
Keras (R10, SM, Adam)        & Rob              & 90.0 & 90.9 \\
Keras (R100, R10, SM, Adam)  & Rob              & 90.6 & 91.4 \\
Keras (R1000, R100, SM, Adam)& Rob              & 90.6 & 91.4 \\
\bottomrule
\end{tabular}
}
\end{table}

Note that there is only a small difference in accuracy between 12 and 30 double planes.
Most clusters originate in the front part of the detector and have less than 10 hits, i.e., less than 5 double planes.
Thus, there are only a few clusters where the reaction products leave the shorter detector.

No clear winner can be identified, as most models achieve a balanced accuracy score of 91\%.
For the application in an experiment, the method with the fastest prediction function should be used.
The Random Forest Classifier might have the lead here, as unscaled data can be used directly and the trained model can be transpiled to \texttt{C} code with sklearn-porter \cite{skpodamo}.
The Keras models are also small enough to not require GPUs and could be incorporated into our \texttt{C++} analysis framework.
Classifying clusters was substantially slower with the model Auto-Sklearn produced.

Not all features contribute to the classification process.
For the Random Forest and Ada Boost Classifiers, the influence of the different features and the probability distributions are shown in \cref{f:clusterreco}.
Note that the energy from time-of-flight $E_{ToF}$ has by far the most impact, followed by the time $T$ and the deposited energy $E_{dep}$.
The energy from time of flight is also the quantity used in physics analyses after the reconstruction process.
Surprisingly, the size of the cluster $N$ as well as its direction $E_M$ play little to no role for both models, the timespan $\Delta T$ is only somewhat relevant for the Random Forest Classifier.
This is similar to the original procedure, where the clusters were ranked by minimal $R$-Value.

\subsection{Cluster selection}
Cluster classification alone is not viable for multiplicity reconstruction.
For example, a BAC of 90\% for classification results in a multiplicity reconstruction BAC of 46\% for $N_{PN}$ and 56\% for $N_{PH}$.

\begin{figure}
\centering
\includegraphics[width=\columnwidth]{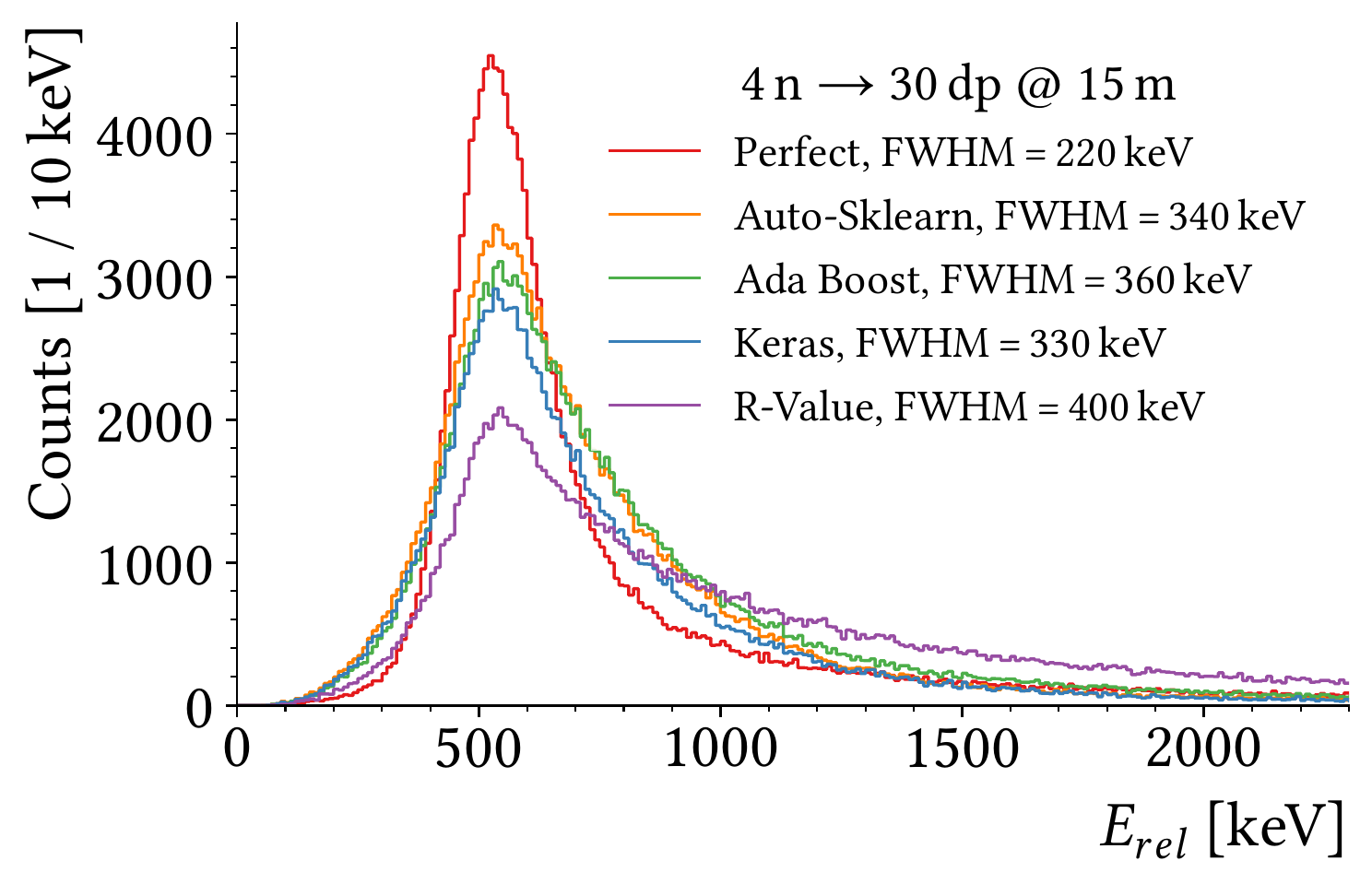}
\caption{Relative energy spectra for 4 incoming neutrons on 30 double planes at \SI{15}{m} distance. The correct multiplicity was provided for each reconstructed event to only study the differences between the different cluster selection methods.
As the peaks are asymmetric with a strong right tail, the \emph{Full Width at Half Max} (FWHM) was measured instead of fitting a Gaussian distribution.
Note that the $R$-Value method shows the worst peak shape.
Auto-Sklearn, Ada Boost, and Keras(100, SM) produce significantly better peak shapes, where the peak created with Keras has the advantage in FWHM but lower volume.
None of the three methods come close to the optimal result.
\label{f:erel}}
\end{figure}

Thus, instead of binary classification, the clusters with the highest probabilities are chosen based on the multiplicity determined beforehand with the methods described in \cref{s:multiplicity}.
With this method, an acceptable first interaction point reconstruction can be achieved.
This can be quantified by the \emph{Full Width at Half Max} of the resulting $E_{rel}$ spectra as defined in \cref{e:erel}.
In \cref{f:erel}, an example is shown where the different cluster ranking models were supplied with the same multiplicity.
The original $R$-Value based method has a strong right tail, which results in a wider peak.
While both Ada Boost and Keras fall short of an optimal reconstruction, they produce significantly narrower and higher peaks.
This can significantly improve experimental sensitivity and thus lead to improved results.

\section{Summary and Outlook}
In this work, we have presented the New Large Area Neutron Detector NeuLAND and the associated challenges in data reconstruction.
These mostly stem from the vastly different interactions neutrons can undergo in the material and the resulting energy deposition patterns.

Different approaches to obtain the multiplicity and the primary interaction points have been investigated.

For the multiplicity reconstruction, many feature-scaler-model combinations were tested.
The classification models from the scikit-learn library and simple Keras-based models only slightly outperform the classical methods.
In addition, the use of more features, e.g., data from all detector elements instead of consolidated quantities, only marginally increases performance.
Thus, for multiplicity reconstruction, we currently recommend using the Bayesian method due to its simplicity, ease of use, and speed, or Keras models for slightly increased accuracy if a balanced approach between all multiplicities is desired.
The calorimetric method shows similar performance; however, its training phase requires optimization of fit parameters.
This can be an advantage if the goal of the experiment requires overweighting or suppressing specific misidentified neutron numbers.

For the reconstruction of the first interaction points in the detector, both scikit-learn- and Keras-based models perform significantly better than ranking by $R$-Value, however, they still fall short of what a hypothetical ideal reconstruction could achieve.
These models require a well-tuned simulation that matches the experimental data and thus exhibits the same behavior for the features that enter the neural network.

It seems prudent to further investigate deep-learning-based approaches.
The Keras-based models presented here are tiny compared to typical neural networks with multiple layers and thousands of nodes, but run fast on pure CPU systems, and deliver good performance in both multiplicity and first interaction point reconstruction.
So far, the reconstruction process has been separated into these two steps, and does not, for example, take advantage of the intercorrelation between the clusters in an event.
A well-crafted convolutional neural network could take the 3D-pixel-image of an event, and directly return the first interaction points.
Alternatively, a Long short-term memory (LSTM) network could be designed to take the cluster intercorrelations into account.
With the rise of \emph{AutoML} methods, further developments could also focus on improving and validating the inputs and leverage the work that others have invested in optimizing model construction.

A detailed comparison of simulated to experimental data is necessary for the more complex models.
Some features like the number of hits and the total deposited energy depend only slightly on the physics lists used, and their distributions can be easily compared to the actual experimental distributions.
For models that analyze individual patterns within the detector, e.g., the 3D- or LTSM networks, a high-quality emulation of the experimental data is vital, as the description of interactions in the detector can differ substantially between physics implementations \cite{Douma2021}.
However, new models can be developed in parallel to this task, and then retrained based on the results.

\appendix
\section{Acknowledgments}
Supported by the BMBF (05P2015PKFNA, 05P19PKFNA) and the GSI (KZILGE1416).

\bibliography{neuland-reco}

\begin{thebibliography}{10}
\expandafter\ifx\csname url\endcsname\relax
  \def\url#1{\texttt{#1}}\fi
\expandafter\ifx\csname urlprefix\endcsname\relax\def\urlprefix{URL }\fi
\expandafter\ifx\csname href\endcsname\relax
  \def\href#1#2{#2} \def\path#1{#1}\fi

\bibitem{FAIR}
\href{https://fair-center.eu/}{{FAIR - Facility for Antiproton and Ion
  Research}}, accessed on 2021-02-15.
\newline\urlprefix\url{https://fair-center.eu/}

\bibitem{Aumann2007}
T.~Aumann, {Prospects of nuclear structure at the future GSI accelerators},
  Progress in Particle and Nuclear Physics 59 (2007) 3--21.
\newblock \href {https://doi.org/10.1016/j.ppnp.2006.12.018}
  {\path{doi:10.1016/j.ppnp.2006.12.018}}.

\bibitem{Geissel2003}
H.~Geissel, H.~Weick, M.~Winkler, G.~Münzenberg, V.~Chichkine, et~al., {The
  Super-FRS project at GSI}, Nuclear Instruments and Methods in Physics
  Research Section B: Beam Interactions with Materials and Atoms 204 (2003)
  71--85.
\newblock \href {https://doi.org/10.1016/S0168-583X(02)01893-1}
  {\path{doi:10.1016/S0168-583X(02)01893-1}}.

\bibitem{Aumann2005}
T.~Aumann, Reactions with fast radioactive beams of neutron-rich nuclei, The
  European Physical Journal A - Hadrons and Nuclei 26~(3) (2005) 441--478.
\newblock \href {https://doi.org/10.1140/epja/i2005-10173-4}
  {\path{doi:10.1140/epja/i2005-10173-4}}.

\bibitem{Baumann2012}
T.~Baumann, A.~Spyrou, M.~Thoennessen, {Nuclear structure experiments along the
  neutron drip line}, Reports on Progress in Physics 75 (2012) 036301.
\newblock \href {https://doi.org/10.1088/0034-4885/75/3/036301}
  {\path{doi:10.1088/0034-4885/75/3/036301}}.

\bibitem{NeuLANDpaper}
K.~Boretzky, I.~Gašparić, M.~Heil, J.~Mayer, A.~Heinz, et~al., {NeuLAND:The
  High-Resolution Neutron Time-of-Flight Spectrometer for R³B at FAIR}, to be
  published.

\bibitem{NeulandTDR}
R3B-Collaboration, \href{https://edms.cern.ch/document/1865739}{{Technical
  Report for the Design, Construction and Commissioning of NeuLAND: The
  High-Resolution Neutron Time-of-Flight Spectrometer for R$^3$B}} (2011).
\newline\urlprefix\url{https://edms.cern.ch/document/1865739}

\bibitem{Bertini2011}
D.~Bertini, {R3BRoot, simulation and analysis framework for the R3B experiment
  at FAIR}, J. Phys. Conf. Ser. 331 (2011) 032036.
\newblock \href {https://doi.org/10.1088/1742-6596/331/3/032036}
  {\path{doi:10.1088/1742-6596/331/3/032036}}.

\bibitem{Al-Turany2012}
M.~Al-Turany, D.~Bertini, R.~Karabowicz, D.~Kresan, P.~Malzacher, et~al., {The
  FairRoot framework}, J. Phys. Conf. Ser. 396 (2012) 022001.
\newblock \href {https://doi.org/10.1088/1742-6596/396/2/022001}
  {\path{doi:10.1088/1742-6596/396/2/022001}}.

\bibitem{Antcheva2009}
I.~Antcheva, M.~Ballintijn, B.~Bellenot, M.~Biskup, R.~Brun, et~al., {ROOT - A
  C++ framework for petabyte data storage, statistical analysis and
  visualization}, Comput. Phys. Commun. 180 (2009) 2499--2512.
\newblock \href {https://doi.org/10.1016/j.cpc.2009.08.005}
  {\path{doi:10.1016/j.cpc.2009.08.005}}.

\bibitem{hdf5}
{The HDF Group}, \href{hdfgroup.org/hdf5}{{Hierarchical Data Format, version
  5}} (1997-2021).
\newline\urlprefix\url{hdfgroup.org/hdf5}

\bibitem{mckinney-proc-scipy-2010}
W.~McKinney, {Data Structures for Statistical Computing in Python}, Proceedings
  of the 9th Python in Science Conference (2010) 56 -- 61\href
  {https://doi.org/10.25080/Majora-92bf1922-00a}
  {\path{doi:10.25080/Majora-92bf1922-00a}}.

\bibitem{reback2020pandas}
{The Pandas Development Team}, Pandas (2020).
\newblock \href {https://doi.org/10.5281/zenodo.3509134}
  {\path{doi:10.5281/zenodo.3509134}}.

\bibitem{protobuf}
{Google}, \href{developers.google.com/protocol-buffers}{{Protocol Buffers -
  Google's data interchange format}} (2021).
\newline\urlprefix\url{developers.google.com/protocol-buffers}

\bibitem{parquet}
{Apache Software Foundation}, \href{parquet.apache.org}{{Apache Parquet}}
  (2021).
\newline\urlprefix\url{parquet.apache.org}

\bibitem{Agostinelli2003}
S.~Agostinelli, J.~Allison, K.~Amako, J.~Apostolakis, H.~Araujo, et~al.,
  {GEANT4 - a simulation toolkit}, Nucl. Instrum. Methods A 506 (2003)
  250--303.
\newblock \href {https://doi.org/10.1016/S0168-9002(03)01368-8}
  {\path{doi:10.1016/S0168-9002(03)01368-8}}.

\bibitem{Kahlbow2018}
J.~Kahlbow, K.~Boretzky, N.~Achouri, D.~Ahn, H.~A. Falou, et~al., {Experimental
  campaign using the NeuLAND demonstrator at SAMURAI}, GSI-FAIR Scientific
  Report 2017 (2018) 151--155\href {https://doi.org/10.15120/GSI-2017-01856}
  {\path{doi:10.15120/GSI-2017-01856}}.

\bibitem{fairsoft}
{FairRootGroup}, \href{github.com/FairRootGroup/FairSoft}{{FairSoft}} (2021).
\newline\urlprefix\url{github.com/FairRootGroup/FairSoft}

\bibitem{Douma2021}
C.~Douma, E.~Hoemann, N.~Kalantar-Nayestanaki, J.~Mayer, {Development of a Deep
  Neural Network for the data analysis of the NeuLAND neutron detector}, Nucl.
  Instrum. Methods A 990 (2021) 164951.
\newblock \href {https://doi.org/10.1016/j.nima.2020.164951}
  {\path{doi:10.1016/j.nima.2020.164951}}.

\bibitem{inclxx}
J.~Cugnon, A.~Boudard, J.-C. David, S.~Leray, D.~Mancusi, {The Li\`ege
  Intranuclear Cascade model - Towards a unified description of nuclear
  reactions induced by nucleons and light ions from a few MeV to a few GeV},
  EPJ Web of Conferences 66 (2014) 03021.
\newblock \href {https://doi.org/10.1051/epjconf/20146603021}
  {\path{doi:10.1051/epjconf/20146603021}}.

\bibitem{PhysRevC.96.054602}
J.~L. Rodr\'{\i}guez-S\'anchez, J.-C. David, D.~Mancusi, A.~Boudard, J.~Cugnon,
  S.~Leray, {Improvement of one-nucleon removal and total reaction cross
  sections in the Li\`ege intranuclear-cascade model using
  Hartree-Fock-Bogoliubov calculations}, Phys. Rev. C 96 (2017) 054602.
\newblock \href {https://doi.org/10.1103/PhysRevC.96.054602}
  {\path{doi:10.1103/PhysRevC.96.054602}}.

\bibitem{scikit-learn}
F.~Pedregosa, G.~Varoquaux, A.~Gramfort, V.~Michel, B.~Thirion, et~al.,
  \href{https://dl.acm.org/doi/10.5555/1953048.2078195}{{Scikit-learn: Machine
  Learning in Python}}, Journal of Machine Learning Research 12 (2011)
  2825--2830.
\newline\urlprefix\url{https://dl.acm.org/doi/10.5555/1953048.2078195}

\bibitem{Chollet2015}
F.~Chollet, et~al., {Keras}, \url{https://keras.io} (2015).

\bibitem{harris2020array}
C.~R. Harris, K.~J. Millman, S.~J. van~der Walt, R.~Gommers, et~al., Array
  programming with {NumPy}, Nature 585~(7825) (2020) 357--362.
\newblock \href {https://doi.org/10.1038/s41586-020-2649-2}
  {\path{doi:10.1038/s41586-020-2649-2}}.

\bibitem{tensorflow2015-whitepaper}
M.~Abadi, A.~Agarwal, P.~Barham, E.~Brevdo, Z.~Chen, et~al.,
  \href{tensorflow.org}{{TensorFlow: Large-Scale Machine Learning on
  Heterogeneous Systems}} (2015).
\newline\urlprefix\url{tensorflow.org}

\bibitem{scaler}
\href{scikit-learn.org/stable/modules/preprocessing.html}{Scikit-learn:
  Preprocessing data} (2021).
\newline\urlprefix\url{scikit-learn.org/stable/modules/preprocessing.html}

\bibitem{NIPS2015_5872}
M.~Feurer, A.~Klein, K.~Eggensperger, J.~Springenberg, M.~Blum, F.~Hutter,
  \href{http://papers.nips.cc/paper/5872-efficient-and-robust-automated-machine-learning.pdf}{{Efficient
  and Robust Automated Machine Learning}}, in: C.~Cortes, N.~D. Lawrence, D.~D.
  Lee, M.~Sugiyama, R.~Garnett (Eds.), Advances in Neural Information
  Processing Systems 28, 2015, pp. 2962--2970.
\newline\urlprefix\url{http://papers.nips.cc/paper/5872-efficient-and-robust-automated-machine-learning.pdf}

\bibitem{skpodamo}
D.~Morawiec, \href{https://github.com/nok/sklearn-porter}{{sklearn-porter}},
  transpile trained scikit-learn estimators to C, Java, JavaScript and others.
\newline\urlprefix\url{https://github.com/nok/sklearn-porter}

\end{thebibliography}
\end{document}